\documentclass[twocolumn,nofootinbib,amsmath,prd,aps,superscriptaddress,tightenlines,preprintnumbers,nobalancelastpage]{revtex4-2}

\usepackage{slashed}
\usepackage{color, verbatim}
\usepackage{latexsym}
\usepackage{amsmath}
\usepackage{graphicx}
\usepackage{amssymb}
\usepackage{graphicx}
\usepackage{bm}
\usepackage{adjustbox}
\usepackage{graphicx}
\usepackage{latexsym}
\usepackage{amsmath}
\usepackage{mathrsfs}
\usepackage{enumerate}
\usepackage{adjustbox}
\usepackage{amssymb}
\usepackage{slashed}
\usepackage{graphicx}    
\usepackage{caption}     
\usepackage{subcaption}    
\usepackage{wrapfig}     
\usepackage{physics}
\usepackage{lipsum}
\usepackage{tikz}
\usetikzlibrary{shapes.geometric}
\usepackage{fontawesome}
\def\link{https://github.com/pabloql/counting-axions-iaxo}

\def\bea{\begin{eqnarray}}
\def\eea{\end{eqnarray}}

\usepackage[normalem]{ulem}
\RequirePackage[colorlinks=true,urlcolor=blue,anchorcolor=blue,citecolor=blue,filecolor=blue,linkcolor=blue,menucolor=blue,linktocpage=true,pdfproducer=medialab,pagebackref=true]{hyperref}
\hypersetup{
    colorlinks = true,
    linkcolor = blueCERN,
    anchorcolor = blueCERN,
    citecolor = blueCERN,
    filecolor = blueCERN,
    urlcolor = blueCERN
    }

\usepackage{orcidlink}
\definecolor{blueCERN}{HTML}{0033A0}
\usepackage[capitalise]{cleveref}
\usepackage{csquotes}
\definecolor{mygreen}{rgb}{0.0, 0.5, 0.0}

\usepackage[capitalise]{cleveref}
\usepackage[font=small,labelfont=bf,justification=centerlast]{caption}
\usepackage{mathtools}

\def\eg{\textit{e.g.}}
\def\ie{\textit{i.e.}}

\newcommand{\Lag}{{\mathscr{L}}}

\definecolor{diamondpink}{HTML}{E73F74}
\definecolor{myyellow}{HTML}{FECC5C}
\definecolor{myred}{HTML}{F03B20}
\newcommand{\reddiamond}{$\mathrel{\ooalign{\textcolor{diamondpink}{$\blacklozenge$}\cr\hidewidth$\lozenge$\hidewidth}}$}
\newcommand{\pinkcross}{\tikz[baseline=-0.7ex]{
  \draw[line width=2.6pt, black] (-3.3pt,-3.3pt) -- (3.3pt,3.3pt);
  \draw[line width=2.6pt, black] (-3.3pt,3.3pt) -- (3.3pt,-3.3pt);
  \draw[line width=1.9pt, diamondpink] (-3pt,-3pt) -- (3pt,3pt);
  \draw[line width=1.9pt, diamondpink] (-3pt,3pt) -- (3pt,-3pt);}}

\newcommand{\redcross}{\tikz[baseline=-0.7ex]{
  \draw[line width=2.6pt, black] (-3.3pt,-3.3pt) -- (3.3pt,3.3pt);
  \draw[line width=2.6pt, black] (-3.3pt,3.3pt) -- (3.3pt,-3.3pt);
  \draw[line width=1.9pt, myred] (-3pt,-3pt) -- (3pt,3pt);
  \draw[line width=1.9pt, myred] (-3pt,3pt) -- (3pt,-3pt);}}
  
\newcommand{\yellowcross}{\tikz[baseline=-0.7ex]{
  \draw[line width=2.6pt, black] (-3.3pt,-3.3pt) -- (3.3pt,3.3pt);
  \draw[line width=2.6pt, black] (-3.3pt,3.3pt) -- (3.3pt,-3.3pt);
  \draw[line width=1.9pt, myyellow] (-3pt,-3pt) -- (3pt,3pt);
  \draw[line width=1.9pt, myyellow] (-3pt,3pt) -- (3pt,-3pt);}}
\newcommand{\pinkcircle}{\tikz[baseline=-0.7ex]{
  \filldraw[fill=diamondpink, draw=black, line width=0.5pt] (0,0) circle (3pt);}}
\newcommand{\pinktriangle}{\tikz[baseline=-0.7ex]{
  \filldraw[fill=diamondpink, draw=black, line width=0.5pt]
    (0,3.5pt) -- (3pt,-2pt) -- (-3pt,-2pt) -- cycle;}}

\definecolor{brightpink}{rgb}{1.0, 0.0, 0.5}

\newcommand{\sinc}{\text{sinc}}
\newcommand{\ag}{a_\gamma}

\usepackage{booktabs}

\begin{document}

\newcount\hour \newcount\minute
\hour=\time \divide \hour by 60
\minute=\time
\count99=\hour \multiply \count99 by -60 \advance \minute by \count99
\newcommand{\mydate}{\ \today \ - \number\hour :00}

\preprint{}

\title{\Large Counting axions with IAXO } 

\author{Benjamín Grinstein\orcidlink{0000-0003-2447-4756}}
\email{bgrinstein@ucsd.edu}
\affiliation{Department of Physics, University of California, San Diego, La Jolla, CA 92093, USA}

\author{Carlos Miró\orcidlink{0000-0003-0336-9025}}
\email{carlos.miroarenas@to.infn.it}
\affiliation{INFN, Sezione di Torino, Via Pietro Giuria 1, I-10125 Turin, Italy}

\author{Pablo Quílez Lasanta\orcidlink{0000-0002-4327-2706}}
\email{pablo.quilez@cern.ch}
\affiliation{Department of Physics, University of California, San Diego, La Jolla, CA 92093, USA}
\affiliation{Theoretical Physics Department, CERN, 1 Esplanade des Particules, CH-1211 Geneva 23, Switzerland}

\preprint{CERN-TH/2026-118}

\begin{abstract}
The existence of multiple axion species is a generic prediction of a number of extensions of the Standard Model. If more than one axion couples to photons, their combined signal in helioscope experiments may mimic that of a single axion with different parameters. This raises a fundamental question: if a next-generation helioscope such as IAXO detected a signal, would we be able to disentangle whether it originated from one or multiple axions? To answer this question, we first recast current CAST bounds and derive IAXO/IAXO+ projections in the two-axion parameter space, identifying the regions where a signal could be observed. Then, we analyze the spectral signatures of \emph{axion flavor oscillations} in both the quasi-degenerate and hierarchical mass regimes, and point out where IAXO can discriminate a two-axion signal from the single-axion hypothesis given the expected energy resolutions of the detector. Finally, we show that these results extend to a broad class of $N$-axion systems.
\end{abstract}

\maketitle


\newpage

\section{Introduction}
\label{sec:introduction}
Axions and axion-like particles (ALPs) are pseudo-Goldstone bosons
arising from spontaneously broken global symmetries in a wide variety
of extensions of the Standard Model (SM). The most prominent example
is the QCD
axion~\cite{Peccei:1977hh,Peccei:1977ur,Weinberg:1977ma,Wilczek:1977pj},
which dynamically solves the strong CP problem, while ALPs are also
common predictions of string
theory~\cite{Witten:1984dg,Arvanitaki:2009fg,Cicoli:2013ana}  and
extra-dimensional theories~\cite{Dienes:1999gw,deGiorgi:2024elx}, among other
examples. In addition, ALPs could
account for the dark matter component of the
Universe~\cite{Abbott:1982af,Dine:1982ah,Preskill:1982cy}. This broad theoretical motivation has driven an ambitious
experimental program searching for these particles~\cite{Irastorza:2018dyq,Adams:2022pbo,DiLuzio:2020wdo,Sikivie:2020zpn}. 

Crucially, many of these theoretical frameworks predict not one but multiple axion-like fields (hereafter simply referred to as axions). String theory and extra dimensions generically give rise to a whole spectrum of such particles, the so-called \emph{axiverse}~\cite{Arvanitaki:2009fg,Dienes:1999gw}. Realizations of the axiverse with characteristic axion mass and coupling patterns have been explored in the type IIB~\cite{Cicoli:2012sz, Demirtas:2018akl, Gendler:2023kjt, Demirtas:2021gsq, Sheridan:2024vtt}, F-theory~\cite{Fallon:2025lvn}, and heterotic~\cite{Benabou:2026jtv} string compactifications. Even the strong CP problem itself can be solved by multiple axions simultaneously~\cite{Gavela:2023tzu}. More broadly, any scalar singlet in Nature would generically mix with the axion, naturally giving rise to a multi-axion scenario.
Despite the strong theoretical motivation for the existence of multiple axions, the vast majority of experimental searches are designed and interpreted within the single-axion paradigm. Only recently have some works begun to explore the phenomenological implications of multiple axion species in laboratory experiments~\cite{Chadha-Day:2021uyt,Gavela:2023tzu,deGiorgi:2025ldc}, astrophysical environments~\cite{Chadha-Day:2023wub}, and cosmological settings related to dark matter production~\cite{Takahashi:2015waa,Ho:2018qur,Cyncynates:2021xzw,Takahashi:2019pqf,Dunsky:2025sgz}. A central finding is that the mismatch between the interaction and mass bases leads to \emph{axion flavor oscillations}, a quantum mechanical phenomenon closely analogous to neutrino flavor oscillations, in which different axion mass eigenstates interfere as they propagate. These oscillations can qualitatively alter the expected signals in axion experiments with respect to the single-axion case.

This naturally raises the question that motivates the present work: \emph{if a next-generation axion experiment detects a signal, can we disentangle whether it originates from one or multiple axions?}

A first candidate among laboratory-based searches would be light-shining-through-wall experiments such as ALPS II~\cite{ALPSII:2025eri}, but their monochromatic laser provides essentially a single measurement, insufficient to resolve whether a signal comes from one or several axions. The continuous spectrum of solar axions, by contrast, amounts to many measurements at once (one per photon energy bin), and this spectral information makes helioscopes a promising candidate to address this question.

Helioscopes search for solar axions by pointing an X-ray detector at the Sun, converting possible incoming ultrarelativistic axions ($\sim$ keV) into photons within a strong magnetic field \cite{Sikivie:1983ip,vanBibber:1988ge}. The CAST experiment~\cite{CAST:2017uph,CAST:2024eil} has set the strongest bound on the axion-photon coupling using this technique, and its successor IAXO~\cite{IAXO:2019mpb} is expected to improve this sensitivity by more than an order of magnitude. A well-known feature of helioscopes is their insensitivity to the axion mass in vast regions of the parameter space: the conversion probability is essentially mass-independent as long as the axion-photon oscillation length exceeds the magnet length. However, as recently shown in Ref.~\cite{Dafni:2018tvj}, for axion masses in the $m_a \sim 10^{-2}$~eV ballpark, the loss of coherence along the magnet imprints characteristic oscillations in the X-ray photon energy spectrum, allowing a measurement of the axion mass from the photon spectral information.

In a two-axion system, an additional oscillatory term shapes the
photon spectrum that is governed by the squared mass difference
$\Delta m_{21}^2 \equiv m_2^2 - m_1^2$ between the two mass
eigenstates. The physical origin is the same as in neutrino
oscillations: the two mass eigenstates, produced coherently through
their common coupling to photons, accumulate different phases as they
travel from the Sun to the detector. This interference produces a
cosine modulation in the photon energy spectrum  which is most
pronounced for mass differences $\sqrt{\Delta m_{21}^2} \sim
10^{-7}$~eV, where the typical oscillation length for keV axions is comparable to the
Earth-Sun distance. 
Importantly, the resulting oscillatory pattern is qualitatively distinct from the single-axion case, making it possible in principle to tell them apart.
Beyond establishing the presence of multiple axions, a measurement of
$\Delta m_{21}^2$ would provide a handle on the mass scale of the
axion system. This could be crucial for uncovering the structure of
the underlying multi-axion landscape, for checking whether the QCD axion sum rule~\cite{Gavela:2023tzu}\footnote{See Ref.~\cite{FernandezNavarro:2026cyu} for modifications of the QCD axion sum rule~\cite{Gavela:2023tzu} under more exotic assumptions.} is satisfied and thus diagnosing whether the multi-axion system solves the strong CP problem, and importantly, for elucidating whether the detected particles could
constitute the dark matter. This would then guide the design of targeted haloscope searches in the appropriate mass range~\cite{Dobrich:2025oso,BREAD:2021tpx,
Baryakhtar:2018doz,
ADMX:2025vom,Melcon:2018dba}.

\begin{figure}[t]
     \includegraphics[width=1.0\columnwidth, trim={7pt 0 0 0}, clip]{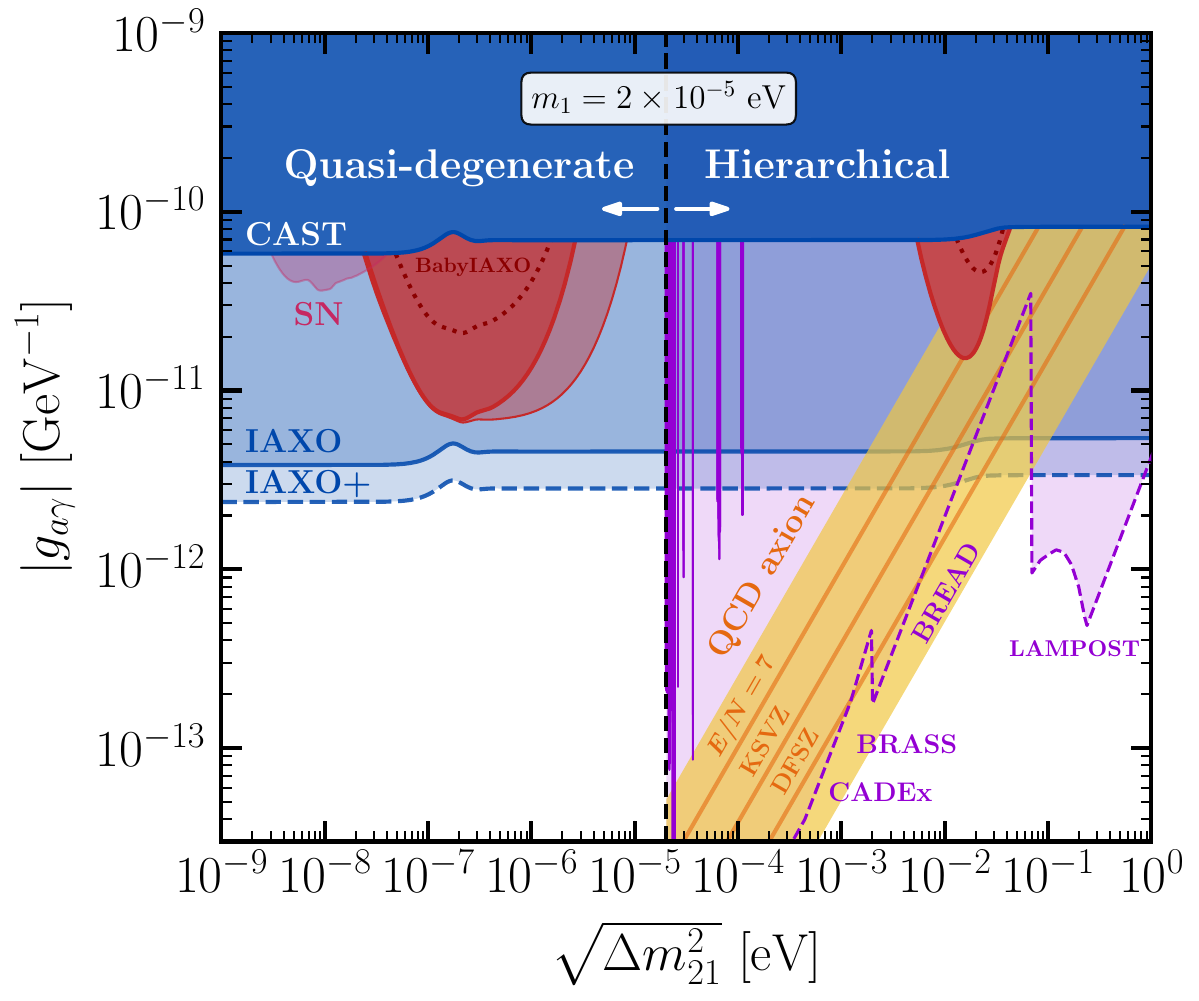}
     \caption{Two-axion discrimination regions (red) in the quasi-degenerate regime $\Delta m_{21}^2\ll m_{1,2}^2$ and hierarchical regime $m_1^2 \ll m_2^2 \simeq \Delta m_{21}^2$. The solid red curves indicate the discrimination reach at IAXO, while the dotted lines assume the BabyIAXO configuration~\cite{IAXO:2020wwp}. The small pink region on the left illustrates the discrimination region derived from the same analysis using supernova axions. CAST exclusion limits at 95\% CL are shown in blue, together with IAXO/IAXO+ projections in light blue. The window for \emph{preferred} QCD axion models is shown in yellow, see Ref.~\cite{DiLuzio:2017pfr}. Finally, since in the hierarchical mass regime, the mass splitting between the two axions is essentially given by the mass of the heaviest axion, we can further include haloscope searches (violet) and projections (light violet) \cite{AxionLimits}. \hfill\,}
     \label{fig:SummaryPlot}
\end{figure}

In this work, we first recast CAST exclusion limits and derive
IAXO/IAXO+ projections in the two-axion parameter space, delimiting
the regions where a two-axion system could produce an observable
signal. Next, we proceed to quantify the ability of IAXO/IAXO+ to
discriminate a two-axion signal from the single-axion hypothesis. To
this end, we carry out a likelihood-based statistical analysis of the photon energy spectrum under the assumption of Asimov data~\cite{Cowan:2010js}, which allows us to identify the regions of the parameter space where the single-axion hypothesis can be rejected in favor of the two-axion hypothesis at least with 3$\sigma$ significance, as shown in red in \cref{fig:SummaryPlot}. We also extend our analysis to the case of supernova (SN) axions and explore whether it would allow IAXO to probe a complementary region of the parameter space if a nearby SN explosion occurs during the experiment's lifetime, as indicated in \cref{fig:SummaryPlot}. 
Finally, we address the generalization to $N$ axions and show that, under certain assumptions, the $N$-axion scenario reduces to the two-axion case for all practical purposes in IAXO, extending the applicability of our results.

The paper is organized as follows. In \cref{sec:multiple_axions_at_iaxo}, we review the helioscope detection principle and derive the axion-photon conversion probability for a two-axion system, discussing the spectral signatures in the quasi-degenerate and hierarchical mass regimes. The details of our statistical analysis based on Asimov data are described in \cref{sec:statistical_analysis}. In \cref{sec:results}, we present the main results of the paper which include the two-axion exclusion bounds and the discrimination reach for IAXO. In \cref{sec:generalization_to_n_axions}, we generalize the formalism to $N$ axions. Finally, we conclude in \cref{sec:discussion_and_outlook}. Additional details on the statistical analysis, the features of CAST/IAXO exclusion limits in the two-axion parameter space, and the study of SN axions are relegated to the appendices. All the code and results of this work are publicly available at \href{\link}{GitHub} \href{\link}{\faGithub}.

\begin{figure*}[t]
\includegraphics[width=0.8\textwidth]{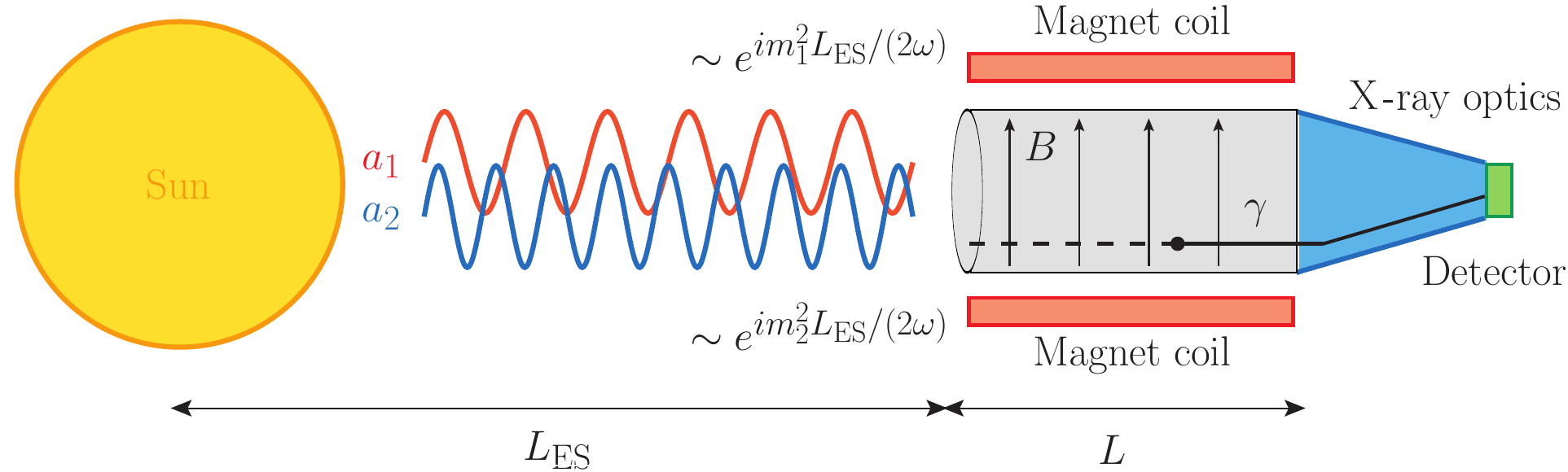}
\caption{Schematic of a two-axion helioscope. The interaction state $a_\gamma$ is produced in the Sun and propagates as a superposition of two mass eigenstates $a_1$ and $a_2$, which accumulate different phase factors $\sim e^{i m_{1,2}^2 L_\mathrm{ES}/(2\omega)}$ over the Earth-Sun distance $L_\mathrm{ES}$. Upon entering the magnet of length $L$ and field strength $B$, the axions convert into X-ray photons. The different accumulated phases generate an interference pattern that imprints oscillatory features on the detected photon energy spectrum, which can be exploited to discriminate the two-axion scenario from the single-axion case.\hfill\,}
\label{fig:helioscope}
\end{figure*} 

\section{Multiple axions at IAXO} 
\label{sec:multiple_axions_at_iaxo}
\subsection{Helioscope detection principle} 
\label{subsec:helioscope_detection_principle}
Helioscope experiments, such as CAST, consist of a long magnetic bore
pointed at the Sun with X-ray detectors at both
ends~\cite{Sikivie:1983ip,Irastorza:2018dyq}. If axions exist and
couple to photons through $g_{a\gamma} a F\tilde F/4$, they will be copiously produced in the solar core
via the Primakoff process $\gamma + q \to a + q$, where $q$ denotes
electrons and nuclei in the solar plasma. The resulting axion flux
travels to Earth and can convert back into X-ray photons in the
transverse magnetic field of the helioscope via the inverse Primakoff
process.  
  The differential axion flux at Earth, as a function of the photon energy $\omega$, can be parametrized as~\cite{CAST:2007jps,Dafni:2018tvj}
\begin{equation}\label{eqn:Primakoff-solar-flux}
\frac{\dd \Phi}{\dd \omega}(\omega) = \Phi_0 \left(\frac{g_{a\gamma}}{10^{-10} \text{ GeV}^{-1}}\right)^2 \frac{\omega^{2.481}}{e^{\omega/1.205}}\,,
\end{equation}
with $\Phi_0 = 6.02 \times 10^{10}$ cm$^{-2}$ s$^{-1}$ keV$^{-1}$, and $\omega$ in units of keV. For the QCD axion, the photon coupling is given by $g_{a\gamma} = \alpha\,(E/N - 1.92)/(2\pi f_a)$, with $f_a$ the axion decay constant, $\alpha$ the fine-structure constant, and $E$ and $N$ the electromagnetic and color anomaly coefficients~\cite{DiLuzio:2020wdo}. The spectrum peaks at $\omega_\mathrm{peak} \simeq 3$~keV and falls off exponentially at higher energies, reflecting the thermal conditions in the solar interior.

The expected number of detected photons is given by
\begin{equation}\label{eqn:Ngamma}
N_{\gamma} = \int \dd\omega\; C(\omega)\; \frac{\dd \Phi}{\dd \omega}(\omega)\; P_{a\to\gamma}(\omega)\,,
\end{equation}
where we have defined $C(\omega) \equiv
S\,t\,\varepsilon_D(\omega)\,\varepsilon_T(\omega)$, with $S$ the
total cross-sectional area of the magnetic bore, $t$ the measurement
time, and $\varepsilon_D$ and $\varepsilon_T$ the detector and
telescope efficiencies, respectively. 

For a single axion of mass $m_a$, the conversion probability in vacuum reads
\begin{equation}\label{eqn:Paxion2photon-single}
P_{a\to\gamma}(\omega) = \left(\frac{g_{a\gamma} B L}{2}\right)^2 \sinc^2\!\left(\frac{m_a^2 L}{4\omega}\right),
\end{equation}
where $B$ is the magnetic field strength, $L$ the magnet length, and
$\sinc\,x \equiv \frac{\sin x}{x}$.\footnote{It is worth noting that
  this function is, however, defined as $\sinc\,x \equiv
  \frac{\sin(\pi x)}{\pi x}$ in languages like \texttt{Python} or
  \texttt{Mathematica}.} This
expression can be derived from the modified Maxwell equations in the
presence of an $a F\widetilde{F}$ coupling (axion electrodynamics),
which in an external magnetic field gives rise to axion-photon
oscillations that can occur coherently over macroscopic
distances~\cite{Raffelt:1987im}.

The analytical dependence of \cref{eqn:Paxion2photon-single} can be understood from the following physical picture~\cite{Redondo:2010dp}. A solar axion entering the helioscope can convert into a photon at any point along the length of the magnet. 
The total conversion amplitude is therefore the coherent sum (integral) of all these individual amplitudes, analogously to the path-integral formulation of quantum mechanics or to the double-slit experiment. Since the axions are ultrarelativistic, their wavenumber is $k_a = (\omega^2 - m_a^2)^{1/2} \simeq \omega - m_a^2/(2\omega)$. Two conversions separated by a distance $\ell$ therefore accumulate a phase difference $\Delta \phi \simeq m_a^2 \ell/(2\omega)$ for a massless photon. When all paths interfere, the average separation $\ell \sim L/2$ sets the characteristic scale, leading to the oscillatory factor $\sin^2(m_a^2 L/4\omega)$. 
The remaining $(m_a^2 L/4\omega)^{-2}$ envelope in $\sinc^2 x = \sin^2 x/x^2$ has a different physical origin, as it is controlled by the effective axion-photon mixing angle $\theta$. In general, the transition probability in a two-state system scales as $\sim \sin^2(2\theta)\sin^2(\Delta\phi)$, where $\theta$ determines the amplitude of the oscillation and $\Delta\phi$ the phase accumulated during propagation. In our case of interest, the interaction $a F\widetilde{F}$ induces an off-diagonal mixing term $\propto g_{a\gamma}\,\omega B$ in the propagation Hamiltonian. Diagonalizing this Hamiltonian gives an effective mixing angle $\sin 2\theta \simeq {2}g_{a\gamma}\omega B/m_a^2$. Importantly, the dependence of the mixing angle on $\omega$ and $m_a$ produces a suppression factor $({2}g_{a\gamma}\omega B/m_a^2)^2{=(g_{a\gamma}BL/2)^2(m_a^2L/4\omega)^{-2}}$ that translates into the $x^{-2}$ envelope. Once it is combined with the oscillatory factor $\sin^2 x$, these two effects together yield the characteristic $\sinc^2\,x$ dependence in the conversion probability.

For small values of the axion mass, $m_a \ll m_{c,\,\mathrm{magnet}} \equiv \sqrt{\omega_\mathrm{peak}/L} \sim 10^{-2}$~eV, the phase difference is negligible and the converted photons interfere constructively. This is the \emph{coherent regime}: $\sinc\ x \simeq 1$, the probability scales as $(g_{a\gamma} BL)^2$ and it becomes completely independent of $m_a$. This explains why the CAST exclusion bounds are insensitive to the axion mass for light axions.

For larger axion masses, $m_a \gtrsim m_{c,\,\mathrm{magnet}}$, the converted photons interfere destructively and the probability drops as $\sin^2 2\theta \simeq (2g_{a\gamma} \omega B/m_a^2)^2$. This is the \emph{incoherent regime}, and explains the loss of sensitivity of CAST at high masses.\footnote{The same reasoning applies to light-shining-through-wall (LSW) experiments: the lower photon energy of the ALPS II laser $\omega \sim 1$~eV~\cite{Ortiz:2020tgs,ALPSII:2025eri} implies $m_{\rm c,\,ALPS\, II} \sim 10^{-4}$~eV. At the other extreme, the proposal of using FASER as an LSW experiment~\cite{Kling:2022ehv,Feng:2022inv} extends its sensitivity to $m_{\rm c,\,FASER} \sim \mathcal{O}(100)$~eV by exploiting TeV-scale photon beams at the LHC.} On the other hand, the oscillatory pattern that the conversion probability imprints on the photon energy spectrum depends on $m_a$, and interestingly would allow IAXO to measure the axion mass in that region from spectral information alone~\cite{Dafni:2018tvj}.


\begin{figure*}[t]
    \includegraphics[width=0.95\textwidth]{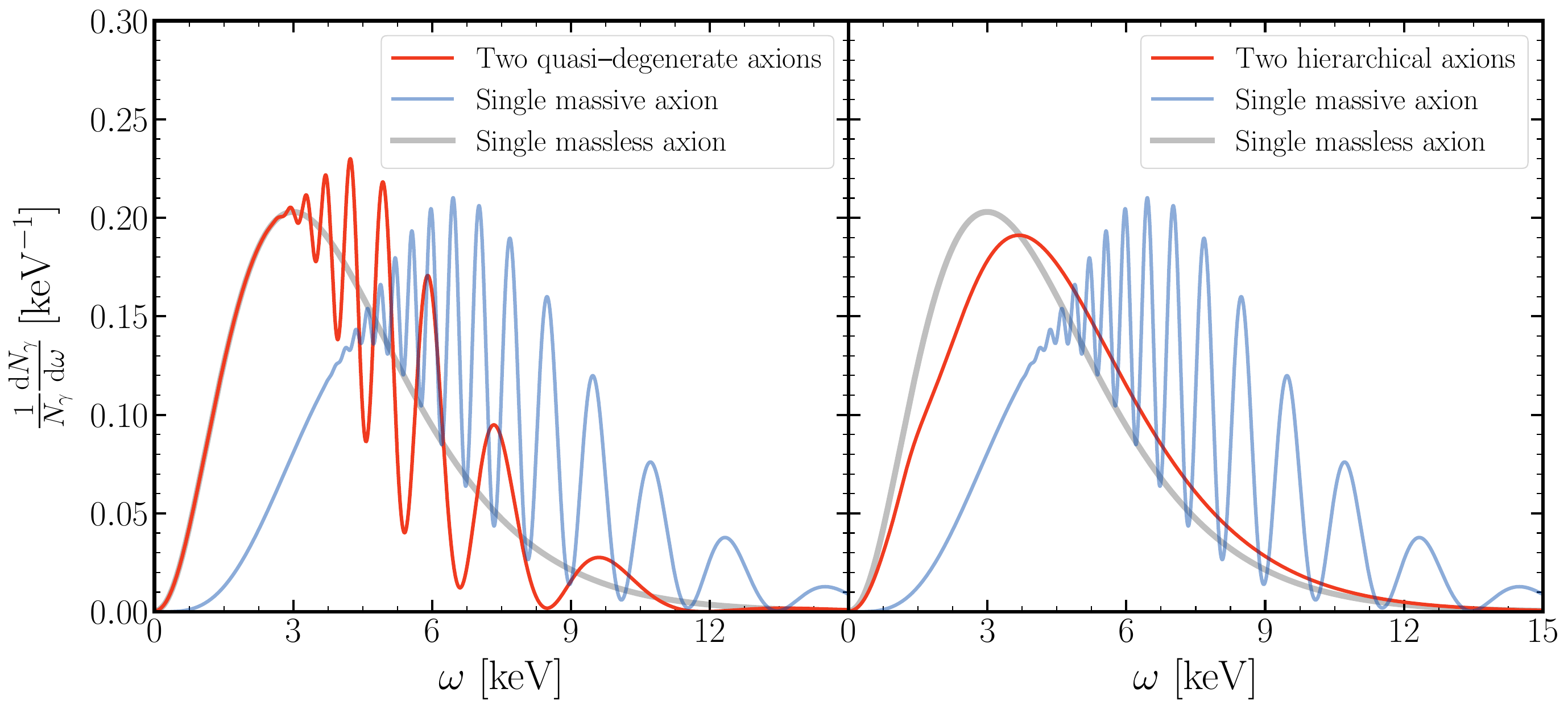}
    \caption{Normalized differential photon spectrum for quasi-degenerate (\emph{left}) and hierarchical (\emph{right}) two-axion scenarios (red), compared to a single massless axion (gray) and a single massive axion (blue). Although both the single massive axion~\cite{Dafni:2018tvj} and the two-axion cases produce an oscillatory pattern that departs from the massless baseline, the spectral shapes are qualitatively different. The benchmark parameters are: $m_a = 10^{-1}$~eV for the massive single-axion theory, $m_2 = 10^{-3}$~eV and $\sqrt{\Delta m_{21}^2} = 7\times 10^{-7}$~eV for the two-axion quasi-degenerate regime, and $\sqrt{\Delta m_{21}^2} \simeq m_2 = 1.5\times 10^{-2}$~eV for the two-axion hierarchical regime. Equal photon couplings  are assumed in all cases ($\varphi = \pi/4$).\hfill\,}
    \label{fig:spectra}
\end{figure*}
\subsection{Axion--photon conversion probability for multiple axions} 
\label{sec:axion_photon_conversion_probability_for_multiple_axions}

We now generalize the single-axion framework of the previous section to the case of two axion-like fields. In the mass basis, the Lagrangian for two axions $\{a_1,a_2\}$ coupled to photons reads
\begin{align}\label{eqn:lagrangian}
\Lag_{\{a_1,\,a_2\}}\supset 
-\frac{1}{2}{m}_1^2 \,a_1^2-\frac{1}{2}{m}_2^2 \,a_2^2 + \Lag_{a\gamma}\,,
\end{align}
where $a_1$ is chosen to be the lightest axion, \ie\ $m_1 < m_2$, and  the axion-photon interaction is given by
\begin{equation}
\Lag_{a\gamma} = -\frac{1}{4}\left(g_1\,a_1 + g_2\,a_2 \right) F_{\mu\nu} \widetilde{F}^{\mu\nu} \equiv - \frac{1}{4}\,g_{a\gamma}\,a_{\gamma}\,F_{\mu\nu} \widetilde{F}^{\mu\nu}\,,
\end{equation}
with $g_{a\gamma} \equiv \sqrt{g_1^2+g_2^2}$, $g_1 \equiv g_{a\gamma}\cos\varphi$, $g_2 \equiv g_{a\gamma}\sin\varphi$, and $a_\gamma \equiv a_1 \cos\varphi + a_2\sin\varphi$ the canonically normalized linear combination that couples to photons. As usual, $F_{\mu\nu} = \partial_\mu A_\nu - \partial_\nu A_\mu$ is the electromagnetic field strength tensor, and $\widetilde{F}_{\mu\nu} \equiv \frac{1}{2}\varepsilon_{\mu\nu\alpha\beta}F^{\alpha\beta}$ its dual. Hereafter, we will often use the shorthand $c_x \equiv \cos x$ and $s_x \equiv \sin x$.

The axion-photon conversion probability in the helioscope can be computed as
\begin{widetext}
    \begin{align}\label{eqn:Paxions2photon-vac}
P_{a_\gamma\,\rightarrow\,\gamma} = \left(\frac{g_{a\gamma} B L}{2}\right)^2\Bigg[&c_\varphi^4\,\sinc^2\left(\frac{m_1^2 L}{4\omega}\right) + s_\varphi^4\,\sinc^2\left(\frac{m_2^2 L}{4\omega}\right) +\frac{1}{2} s_{2\varphi}^2\, \sinc\left(\frac{m_1^2 L}{4\omega}\right) \sinc\left(\frac{m_2^2 L}{4\omega}\right) \cos\left(\frac{\Delta m_{21}^2 L_\mathrm{ES}}{2\omega}\right) \Bigg]\,,
    \end{align}
\end{widetext}
where $\omega$ is the photon energy, $L_\mathrm{ES} \simeq 1$~AU is the Earth-Sun distance, and $\Delta m_{21}^2 \equiv m_2^2 - m_1^2 > 0$ is the squared mass difference (positive by convention).

In close analogy with neutrino oscillations~\cite{Chadha-Day:2021uyt}, $a_\gamma$ is the interaction state produced at the Sun through its coupling to photons, but the propagation from the Sun to the detector is governed by the mass eigenstates $a_1$ and $a_2$, which are the eigenstates of the free Hamiltonian.\footnote{This treatment has known subtleties in the neutrino case, where it can lead to inconsistencies regarding simultaneous energy-momentum conservation~\cite{Beuthe:2001rc,Cohen:2008qb}. A rigorous approach recognizes that the produced state is entangled with the source particle in the Primakoff process $\gamma + q \to a + q$, so that energy and momentum are always conserved. Alternatively, in the quantum field theory approach of Ref.~\cite{Kobach:2017osm}, production and detection are computed simultaneously and the oscillating particle appears only as a virtual exchanged particle. Both treatments reproduce the standard oscillation formula in the suitable limits. The axion case is completely analogous.}
Since $a_1$ and $a_2$ have different masses, they accumulate different phases as they propagate (see helioscope sketch in \cref{fig:helioscope}), causing $a_\gamma$ to partially oscillate into the orthogonal combination $a_\mathrm{dark} \equiv -s_\varphi\, a_1 + c_\varphi\, a_2 $, which does not couple to photons and therefore cannot reconvert in the magnet.

The structure of \cref{eqn:Paxions2photon-vac} reflects this physics. The first two terms describe the independent conversion of each mass eigenstate inside the magnet, weighted by $c_\varphi^4$ and $s_\varphi^4$ (the fourth power arises because $a_\gamma$ has a $c_\varphi^2$ component of $a_1$, whose squared coupling to photons is $g_{a\gamma}^2 c_\varphi^2$, giving $c_\varphi^4$ overall, and analogously for $a_2$), with the same $\sinc^2$ coherence factor discussed in \cref{subsec:helioscope_detection_principle}. 
The third term encodes the $a_\gamma$ disappearance, \ie\, $a_\gamma \to a_\mathrm{dark}$ oscillation. 
The resulting conversion probability cannot exceed that of a single axion with the same total coupling $g_{a\gamma}$: since $g_{a\gamma}^2 = g_1^2 + g_2^2$ is fixed (by the canonical normalization of the interaction state $a_\gamma$), part of $a_\gamma$ oscillates into $a_\mathrm{dark}$ during propagation, thus reducing the signal.\footnote{One can easily show that
$P_{\text{2-axion}}\leq 
(g_{a\gamma} B L/2)^2 \times \max\!\left(\sinc^2\,x_1,\,\sinc^2\,x_2 \right)
\leq (g_{a\gamma} B L/2)^2$,
\ie\ the two-axion conversion probability is always bounded from above by a single-axion result.}
Note also the different factors in the $\sinc$ and cosine arguments: the $\sinc$ phase $m_a^2 L/(4\omega)$ involves the average distance $L/2$ between conversion points inside the magnet (as explained in \cref{subsec:helioscope_detection_principle}), while the cosine phase $\Delta m_{21}^2 L_\mathrm{ES}/(2\omega)$ reflects the fact that each mass eigenstate traverses the entire distance $L_\mathrm{ES}$ before converting, so the full baseline enters the phase difference.

The interference is present even if a single photon is converted. In the path integral language, each axion takes both \emph{paths} through $a_1$ and $a_2$ simultaneously, and the respective converted photons interfere, analogous to a double-slit experiment. No phase coherence between different production events is required, since axions from independent Primakoff processes arrive at the detector at different times and never interfere with each other. A related concern is that different Primakoff events originate at different points inside the solar core, so the propagation distance varies by $\sim R_\mathrm{S}$. Expanding the phase for a production point at the solar limb, $\Delta m_{21}^2 (L_\mathrm{ES} + R_\mathrm{S})/(2\omega)$, the correction is of order $\Delta m^2_{21}R_\mathrm{S}/(2\omega) \sim 5\times10^{-3}$ for $\Delta m^2_{21}\sim (10^{-7}$ eV)$^2$ and $\omega=\omega_{\rm peak}$, so the decoherence is negligible.

The two-axion formula in \cref{eqn:Paxions2photon-vac} reduces to the single-axion case in \cref{eqn:Paxion2photon-single} in three independent limits.
(i)~When $\Delta m_{21}^2 \to 0$, \ie\ $m_2 \to m_1$, the cosine goes to unity and the identity $c_\varphi^4 + s_\varphi^4 + \tfrac{1}{2}s_{2\varphi}^2 = (c_\varphi^2 + s_\varphi^2)^2 = 1$ recovers \cref{eqn:Paxion2photon-single}. If both masses are equal, the mass matrix is proportional to the identity and $\varphi$ is unphysical, so one can always rotate to a mass basis where a single eigenstate carries the full coupling, namely, $\{a_\gamma,\,a_\mathrm{dark}\}$.
(ii)~When $\varphi = 0$ or $\pi/2$, only one mass eigenstate ($a_1$ or $a_2$, respectively) couples to photons and the other is an uncoupled spectator.
(iii)~When the heaviest axion $a_2$ satisfies $m_2 \gg m_{c,\,\mathrm{magnet}}$, the corresponding $\sinc^2$ factor is strongly suppressed, which effectively leads to a single axion with mass $m_a = m_1$ and coupling $g_{a\gamma}^{\text{1-axion}} =g_1= g_{a\gamma}c_{\varphi}$.
\noindent The first two cases illustrate the convenience of defining the total coupling as $g_{a\gamma} \equiv \sqrt{g_1^2 + g_2^2}$: in both limits the probability reduces exactly to the single-axion formula with coupling $g_{a\gamma}^{\text{1-axion}}=g_{a\gamma}$.

Finally, the characteristic length over which the different mass eigenstates accumulate different phases is the Earth-Sun distance $L_\mathrm{ES}$. The critical mass splitting for which the oscillation length becomes comparable to $L_\mathrm{ES}$ is $\sqrt{\Delta m_{21}^2} \sim m_{c,\,\mathrm{Sun}} \equiv \sqrt{\omega_\mathrm{peak}/L_\mathrm{ES}} \sim 10^{-7}$~eV, many orders of magnitude smaller than $m_{c,\,\mathrm{magnet}}$. In summary, nontrivial two-axion features appear in the photon spectrum whenever $m_{1,2} \sim m_{c,\,\mathrm{magnet}}$ (through the $\sinc^2$ terms) or $\sqrt{\Delta m_{21}^2} \sim m_{c,\,\mathrm{Sun}}$ (through the cosine term). We now focus on two limiting regimes of \cref{eqn:Paxions2photon-vac}, illustrated in \cref{fig:spectra}, each dominated by one of the two critical scales and giving rise to spectral features that enable two-axion discrimination.

\subsubsection{Quasi-degenerate regime}\label{subsubsec:quasi_degenerate}

When $m_1 \simeq m_2$ (\ie\ $\Delta m_{21}^2 \ll m_{1,2}^2$), the two $\sinc$ factors in \cref{eqn:Paxions2photon-vac} are approximately equal, $\sinc(m_1^2 L/4\omega) \simeq \sinc(m_2^2 L/4\omega)$, and \cref{eqn:Paxions2photon-vac} reduces to
\begin{multline}\label{eqn:P-quasi-degenerate-general}
P_{a_\gamma\to\gamma} \simeq \left(\frac{g_{a\gamma} BL}{2}\right)^2 \sinc^2\!\left(\frac{m_1^2 L}{4\omega}\right) \\[5pt]
\times \Bigg[c_\varphi^4 + s_\varphi^4 
+ \tfrac{1}{2} s_{2\varphi}^2\, \cos\!\left(\frac{\Delta m_{21}^2 L_\mathrm{ES}}{2\omega}\right)\Bigg]\,.
\end{multline}
Using the identity $c_\varphi^4 + s_\varphi^4 = 1 - \tfrac{1}{2}s_{2\varphi}^2$, this can equivalently be written as a disappearance-type formula, $P \propto 1 - \tfrac{1}{2}s_{2\varphi}^2\left[1 - \cos\left(\Delta m_{21}^2 L_\mathrm{ES}/2\omega\right)\right]$, which will be familiar to the neutrino community. 
If, in addition, both masses lie in the coherent regime ($m_{1,2} \ll m_{c,\,\mathrm{magnet}}$), $\sinc \simeq 1$ and the signal is entirely governed by the cosine modulation. As discussed above, this modulation becomes resolvable when $\sqrt{\Delta m_{21}^2} \sim m_{c,\,\mathrm{Sun}}$, producing oscillations in the photon spectrum qualitatively different from the $\sinc^2$ pattern of a single massive axion. As shown in the left panel of \cref{fig:spectra}, in the quasi-degenerate regime the cosine modulates the spectrum around the massless baseline without shifting it, whereas the single-axion case not only introduces oscillations but also shifts the peak to higher energies, making the two cases distinguishable. For $\sqrt{\Delta m_{21}^2} \ll m_{c,\,\mathrm{Sun}}$ the oscillation is too slow to be observable, while for $\sqrt{\Delta m_{21}^2} \gg m_{c,\,\mathrm{Sun}}$ it is too fast to be resolved by the finite energy resolution $\omega_\mathrm{res}$ of the detector and averages out, leaving a spectrum indistinguishable from a single axion with an effective coupling suppressed by a factor $(1 - \tfrac{1}{2}s_{2\varphi}^2)^{1/2}$.

\subsubsection{Hierarchical regime}\label{subsubsec:hierarchical}

 When one axion is much lighter than the other, $m_1 \ll m_2$, we have $\Delta m_{21}^2 \simeq m_2^2$, and the first $\sinc$ factor satisfies $\sinc^2(m_1^2 L/4\omega) \simeq 1$.  Therefore, \cref{eqn:Paxions2photon-vac} reduces to
\begin{multline}\label{eqn:P-hierarchical}
P_{a_\gamma\to\gamma} \simeq \left(\frac{g_{a\gamma} BL}{2}\right)^2 \Bigg[c_\varphi^4 + s_\varphi^4\,\sinc^2\!\left(\frac{m_2^2 L}{4\omega}\right) \\[5pt]
+ \tfrac{1}{2} s_{2\varphi}^2\, \sinc\!\left(\frac{m_2^2 L}{4\omega}\right) \cos\!\left(\frac{m_2^2 L_\mathrm{ES}}{2\omega}\right)\Bigg]\,.
\end{multline}
In this regime, there are in principle two ranges where the two-axion spectrum develops nontrivial features: $m_2 \sim m_{c,\,\mathrm{magnet}}$, where the $\sinc$ becomes relevant, or $\sqrt{\Delta m_{21}^2} \simeq m_2 \sim m_{c,\,\mathrm{Sun}}$, where the cosine modulation becomes resolvable. In practice, for IAXO only the first case is relevant: the second would require $m_1 \lesssim 10^{-7}$~eV with couplings in the range $g_{a\gamma} \sim 10^{-12}$--$10^{-10}$~GeV$^{-1}$, which is already excluded by NuSTAR and other astrophysical constraints --- see \eg\ Ref.~\cite{AxionLimits}. The phenomenologically viable scenario is therefore $m_2 \sim m_{c,\,\mathrm{magnet}}$, where the cos term averages out and the $\sinc^2$ oscillations dominate. Interestingly, even though both the single massive axion and the hierarchical two-axion case produce $\sinc^2$ oscillations, the spectra are distinguishable: the two-axion probability in \cref{eqn:P-hierarchical} has the structure $c_\varphi^4 + s_\varphi^4\,\sinc^2$, which decouples the overall normalization from the oscillatory part, whereas a single massive axion gives a pure $\sinc^2$ that simultaneously shifts the peak and modulates the spectrum (see right panel of \cref{fig:spectra}). 
 
\section{Statistical analysis}\label{sec:statistical_analysis}
We proceed now to summarize the statistical analysis carried out in this work --- see Appendix~\ref{app:hypothesis-testing} for further details. We rely on a likelihood-based test statistic under the assumption of Asimov data \cite{Cowan:2010js} in order to identify in which regions of the parameter space one would be able to discriminate a two-axion theory from the single-axion hypothesis in the event of a positive signal at next-generation helioscopes. In this sense, it is also important to recast current exclusion limits and project future sensitivity regions in the two-axion parameter space.

To this end, the relevant observable for our analysis is the number of photon counts at each energy, \ie\ an X-ray spectrum such as those shown in \cref{fig:spectra}. Given the high energy resolution of current and future detection technologies, it is useful to perform the usual binned analysis of the photon spectrum in the small bin size limit. This will also allow us to derive approximate analytical formulae to understand the parametric dependence of some of the main results of this work. We have verified that the binned and unbinned statistical analyses yield the same results. 

In the small bin size limit, the likelihood function\footnote{It is often convenient to work with the log-likelihood function: since the logarithm is monotonic, both are maximized by the same parameter values. Thus, likelihood and log-likelihood will be used interchangeably throughout the discussion. Likewise, we have omitted those terms that are independent of the model parameters as they play no role in the maximization and cancel out in the likelihood-ratio test statistic defined below.} constructed from the Poisson probability distribution can be written as
\begin{align}\label{eqn:likelihood}
\log\mathcal{L}(\boldsymbol{\theta}) = \int \dd\omega \Big\{n_\mathrm{obs}(\omega)\log\left[n_\mathrm{exp}(\omega\,;\boldsymbol{\theta})\right]
- n_\mathrm{exp}(\omega\,;\boldsymbol{\theta})\Big\}\,,
\end{align}
where the model parameters are $\boldsymbol{\theta} = (g_{a\gamma},m_2,\Delta m_{21}^2,\varphi)$, and $n_\mathrm{obs/exp}\equiv{\dd N_\mathrm{obs/exp}}/{\dd\omega}$ is the observed/expected differential number of photon counts, \ie\ the number of observed/expected photons per unit energy. The expected photon spectrum depends on the axion parameters through
\begin{equation}\label{eqn:nexp}
n_\mathrm{exp}(\omega\,; \boldsymbol{\theta}) = 
g_{a\gamma}^4\, C(\omega)\, \frac{\dd \widetilde{\Phi}}{\dd \omega}(\omega)\, \widetilde{P}(\omega\,; m_2,\Delta m_{21}^2,\varphi)\,,
\end{equation}
where the solar axion flux $\frac{\dd \Phi}{\dd \omega}(\omega\,; g_{a\gamma})$ and the axion-photon conversion probability $P(\omega\,; g_{a\gamma}, m_2,\Delta m_{21}^2,\varphi)$ are given in \cref{eqn:Primakoff-solar-flux,eqn:Paxions2photon-vac}, respectively, and quantities with a tilde have their $g_{a\gamma}$ dependence factored out, \ie
\begin{align}
\frac{\dd \Phi}{\dd \omega}(\omega\,; g_{a\gamma}) &\equiv g_{a\gamma}^2\; \frac{\dd \widetilde{\Phi}}{\dd \omega}(\omega)\,, \label{eqn:solar-flux-tilde} \\[5pt]
P(\omega\,; g_{a\gamma}, m_2,\Delta m_{21}^2,\varphi) &\equiv g_{a\gamma}^2\; \widetilde{P}(\omega\,; m_2,\Delta m_{21}^2,\varphi)\,, \label{eqn:probability-tilde} \\[5pt]
n_\mathrm{exp}(\omega\,; g_{a\gamma},m_2,\Delta m_{21}^2,\varphi) &\equiv g_{a\gamma}^4\; \widetilde{n}_\mathrm{exp}(\omega\,; m_2,\Delta m_{21}^2,\varphi)\,. \label{eqn:nexp-tilde}
\end{align}

\subsection{Exclusion limits}\label{subsec:cast_recast}
The CAST experiment \cite{CAST:2017uph,CAST:2024eil} has set the strongest limits on the axion-photon coupling $g_{a\gamma}$ among helioscope searches, while IAXO/IAXO+ \cite{IAXO:2019mpb} is expected to improve this bound by more than one order of magnitude in future searches. In the latest data-taking campaign of CAST~\cite{CAST:2024eil}, with improved background rejection from the installed X-ray optics, no photon counts were detected in the signal region, while the expected background was $b = 0.75$. This yields an upper limit of $5.8\times 10^{-11}$ GeV$^{-1}$ at 95\% CL for $m_a \lesssim 0.02$ eV. 

In order to translate this limit into the two-axion parameter space, we can impose that the Poisson probability of detecting zero counts given a total of $N_\mathrm{exp} + b$ expected signal plus background events must satisfy
\begin{equation}\label{eqn:exclusion-limit-condition}
\mathscr{P}[\,0\,|\, N_\mathrm{exp}(g_{a\gamma},m_2,\Delta m_{21}^2,\varphi) + b\,] \geq 1-\alpha\,,
\end{equation}
at a confidence level $\alpha$ ($0<\alpha<1$).

\subsection{Hypothesis discrimination} \label{subsec:two_axion_discrimination}
Our goal is to determine the regions of the parameter space where the single-axion hypothesis can be rejected in favor of the two-axion theory. As explained in \cref{sec:axion_photon_conversion_probability_for_multiple_axions}, the former can be recovered as a limiting case of the two-axion scenario by an appropriate choice of the theory parameters, namely, $\Delta m_{21}^2 = 0$ and/or $\varphi = 0,\,\pi/2$. Consequently, we define our null hypothesis $H_0$ and alternative hypothesis $H_1$ to be
\begin{align}
&H_0: \;\; \Delta m_{21}^2 = 0 \; \,\,\text{ or }\,\, \; \varphi=0,\pi/2 \;\;\implies\; \text{1-axion}\,, \\[5pt]
&H_1: \;\; \Delta m_{21}^2 \neq 0 \; \text{ and } \; \varphi\neq 0,\pi/2 \;\;\implies\; \text{2-axion}\,.
\end{align}
In order to compare the two theories, we construct the likelihood ratio test statistic $q_0$ \cite{Cowan:2010js}:
\begin{equation}\label{eqn:q0}
q_0 = \begin{cases} \;-2\log \Lambda\,, \;\;\quad \Delta \widehat{m}_{21}^2 > 0\\[5pt] \;0\,, \qquad\quad\;\;\;\;\;\; \Delta \widehat{m}_{21}^2 < 0 \end{cases}\,, 
\end{equation}
with
\begin{equation}\label{eqn:likelihood-ratio}
\Lambda = \frac{\mathcal{L}(\widehat{\widehat{g}}_{a\gamma},\widehat{\widehat{m}}_2,0,0)}{\mathcal{L}(\widehat{g}_{a\gamma},\widehat{m}_2,\Delta \widehat{m}_{21}^2,\widehat{\varphi})}\,.
\end{equation}
The parameters $(\widehat{\widehat{g}}_{a\gamma},\widehat{\widehat{m}}_2)$ in the numerator of \cref{eqn:likelihood-ratio} denote the values of $g_{a\gamma}$ and $m_2$ that maximize the likelihood function for the specified $\Delta m_{21}^2 = 0$ and $\varphi = 0$, that is, they are the conditional maximum likelihood estimators (MLE) of $g_{a\gamma}$ and $m_2$ for the null hypothesis $H_0$. On the other hand, the denominator corresponds to the (unconditional) maximum likelihood function, and $(\widehat{g}_{a\gamma},\widehat{m}_2,\Delta \widehat{m}_{21}^2,\widehat{\varphi})$ are their MLE. We should emphasize that the conditional and unconditional MLE do not coincide in general. All in all, the larger the value of $q_0$, the larger the incompatibility between the two hypotheses, that is, the easier to discriminate the two-axion theory against the single-axion theory. 

The likelihood function in \cref{eqn:likelihood} depends on the observed data, which are not yet available for IAXO. Instead, we are interested in estimating the median discrimination significance at IAXO. This motivates the use of the so-called Asimov dataset~\cite{Cowan:2010js}: the two-axion hypothesis $H_1$ is assumed to be the true theory, so that the \emph{observed} data exactly matches the two-axion prediction with no Poisson fluctuations, \ie\ $n_\mathrm{obs}(\omega) = n_\mathrm{exp}(\omega\,; g_{a\gamma},m_2,\Delta m_{21}^2,\varphi)$, and the MLE coincide with the true parameter values, $(\widehat{g}_{a\gamma},\widehat{m}_2,\Delta \widehat{m}_{21}^2,\widehat{\varphi})= ({g}_{a\gamma},{m}_2,\Delta {m}_{21}^2,{\varphi})$. Importantly, under the assumption of Asimov data, the median discrimination significance is simply $n_\sigma = \sqrt{q_0}$.
 Therefore, for a given target significance, we want to compute the minimal coupling $g_{a\gamma}^{\rm dis}$ for which a two-axion signal can be discriminated at $n_\sigma$ sigmas as a function of the remaining model parameters, that is,
\begin{equation}
q_0(g_{a\gamma}^{\rm dis},m_2,\Delta m_{21}^2,\varphi)=n_\sigma^2\,.
\end{equation}
The use of Asimov data is a standard technique for this kind of analysis in the large sample limit, that is assuming a large number of photon counts --- see \eg\ Ref.~\cite{Dafni:2018tvj} ---, which avoids the computational cost of Monte Carlo simulations. The large sample limit is justified \emph{a posteriori} in our analysis since $N_{\rm exp} > 10$ in all the discrimination region.

\section{Results}\label{sec:results} 

\subsection{Exclusion limits for two axions}\label{subsec:exclusion-limits_2-axions}

\begin{table}[t]
\centering
\renewcommand{\arraystretch}{1.25}
\setlength{\tabcolsep}{4pt}
\begin{tabular}{lcccc}
\toprule
 & \textbf{CAST} & \textbf{BabyIAXO} & \textbf{IAXO} & \textbf{IAXO+} \\
\hline
$B$ [T]                      & $9$    & $2$    & $2.5$  & $3.5$  \\
$L$ [m]                      & $9.26$ & $10$   & $20$   & $22$   \\
$S$ [m$^2$]                  & $0.003$& $0.77$ & $2.3$  & $3.9$  \\
$t$ [yr]                     & $0.036$& $1.5$  & $3$    & $5$    \\
$\varepsilon_T$              & $0.6$  & $0.7$  & $0.8$  & $0.8$  \\
$\varepsilon_D$              & $0.3$  & $0.35$ & $0.7$  & $0.7$  \\
\hline
$\omega_\mathrm{res}$ [eV]  & \multicolumn{4}{c}{$10$--$200$}    \\
\bottomrule
\end{tabular}
\caption{Experimental configuration for current and next-generation helioscopes~\cite{IAXO:2020wwp}: magnetic field $B$, length $L$, total aperture area $S$, measurement time $t$, telescope and detector efficiencies $\varepsilon_T$ and $\varepsilon_D$, and range of possible detector energy resolutions $\omega_\mathrm{res}$. For our analysis we set an energy threshold of $100$~eV. The total number of tracking hours for CAST is taken from Ref.~\cite{CAST:2024eil}.\hfill\,}
\label{tab:Helioscopes-configuration}
\end{table}

The relevant experimental parameters of CAST and planned helioscopes such as BabyIAXO, IAXO, and IAXO+ are summarized in \cref{tab:Helioscopes-configuration}. We adopt the IAXO conceptual design values~\cite{IAXO:2019mpb} and, following Ref.~\cite{Dafni:2018tvj}, assume flat efficiency functions and zero background --- the latter justified by the extremely low background rates expected for the IAXO detectors~\cite{IAXO:2019mpb,IAXO:2020wwp}. The finite energy resolution  of the X-ray detectors $\omega_\mathrm{res}$ is modeled by convolving the theoretical spectrum with a Gaussian of width $\omega_\mathrm{res}$, which we vary in the range $10$--$200$~eV to bracket the performance of different detector technologies that could be installed in IAXO (\eg\ Micromegas, metallic magnetic calorimeters, or silicon drift detectors).

As explained in \cref{subsec:cast_recast}, we must translate current CAST bounds and project the sensitivity regions of future helioscopes into the two-axion parameter space. Following \cref{eqn:exclusion-limit-condition}, the \emph{allowed} values of the axion-photon coupling are
\begin{equation}\label{eqn:exclusion-limits}
g_{a\gamma} \leq \left(\frac{-\log(1-\alpha)-b}{\widetilde{N}_\mathrm{exp}(m_2,\Delta m_{21}^2,\varphi)}\right)^{1/4}\,,
\end{equation}
where $\widetilde{N}_\mathrm{exp}$ is computed using the benchmark configurations in \cref{tab:Helioscopes-configuration}, and the background is taken to be $b = 0.75$ for CAST~\cite{CAST:2024eil} and $b = 0$ for IAXO/IAXO+. 

\begin{figure*}[t]
\includegraphics[width=1.0\columnwidth, trim={7pt 0 0 0}, clip]{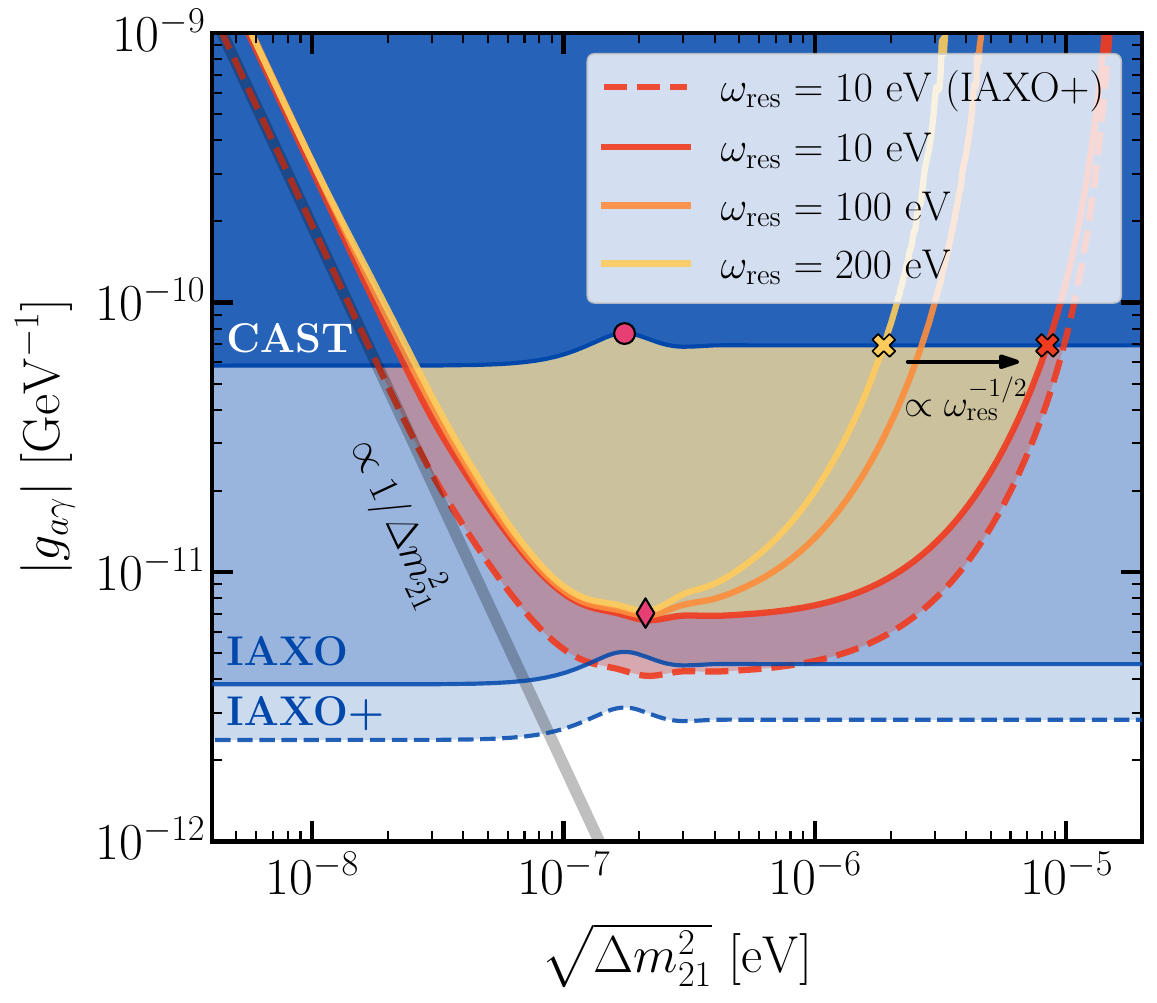}\quad
 \includegraphics[width=1.0\columnwidth, trim={7pt 0 0 0}, clip]{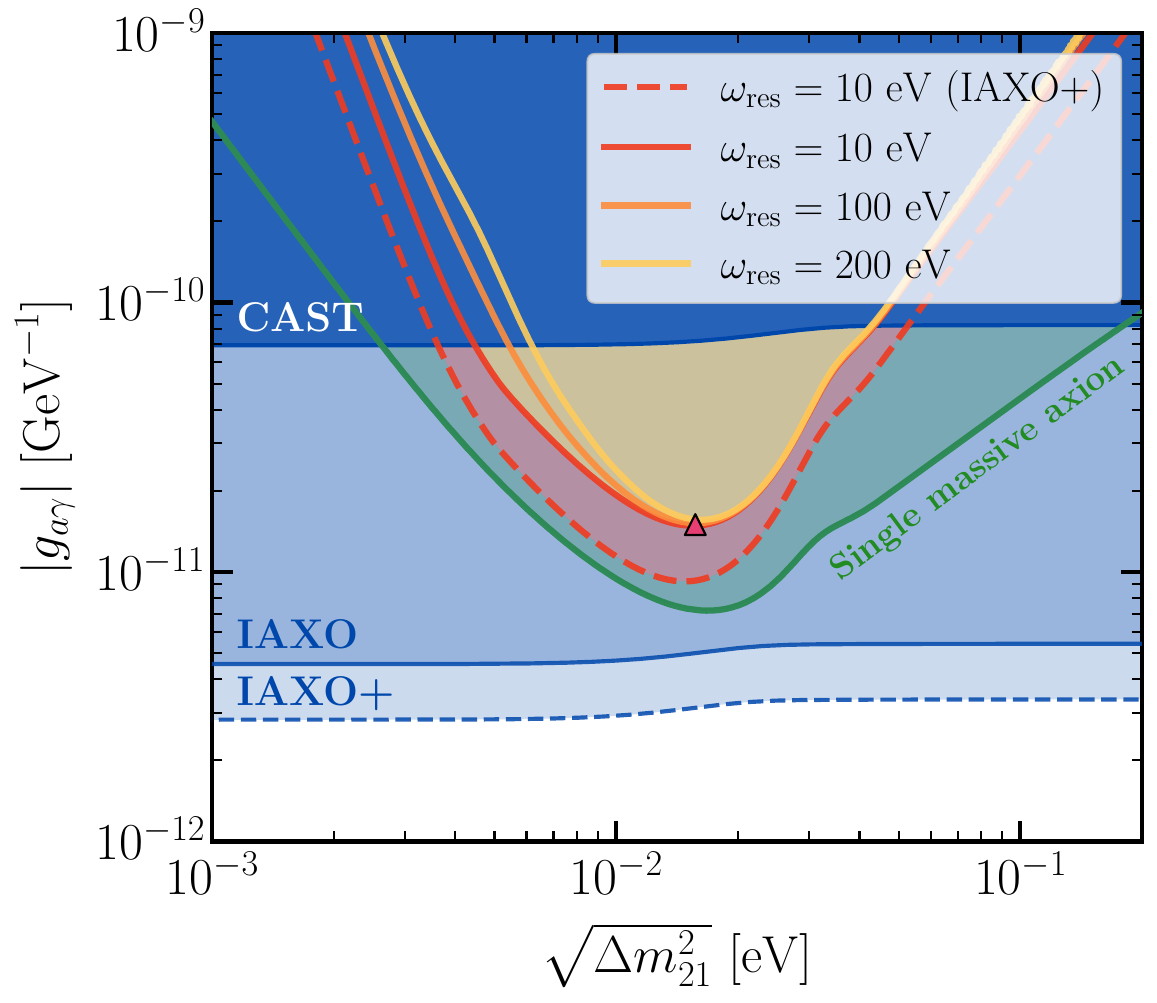}
     \caption{Two-axion discrimination reach (shaded yellow) for equal photon couplings $\varphi = \pi/4$ in the quasi-degenerate regime $\Delta m_{21}^2 \ll m_{1,2}^2$  (\emph{left}) and hierarchical regime $m_1^ 2 \ll m_2^ 2 \simeq {\Delta m_{21}^2}$ (\emph{right}). In both panels, the solid blue curve is the CAST exclusion limit derived in \cref{subsec:exclusion-limits_2-axions}, the light blue band marks IAXO/IAXO+ projections, and the yellow, orange and red lines show $g_{a\gamma}^{\rm dis}$ at $3\sigma$ for energy resolutions $\omega_{\rm res} = 10,\,100,\,200$ eV, respectively (IAXO+ shown as dashed red). \emph{Left:} The marker \protect\reddiamond\ signals the optimal discrimination point, \ie\ the minimum value of $g_{a\gamma}^\mathrm{dis}$, while \scalebox{0.85}{\protect\pinkcircle}\ indicates the mass splitting with the weakest exclusion limit (see details in Appendix \ref{app:details-exclusion-limits}). 
     The maximum mass splitting that can be probed scales as $\propto\omega_{\rm res}^{-1/2}$, as indicated by the horizontal arrow when the resolution is improved from 200 eV (\scalebox{0.85}{\protect\yellowcross}) down to 10 eV (\scalebox{0.85}{\protect\redcross}).
     \emph{Right:} The green line marks where IAXO can measure the single axion mass as computed by Ref.~\cite{Dafni:2018tvj}, and \protect\pinktriangle\ indicates again the optimal discrimination point.
     \hfill\,}
     \label{fig:quasiDegRes}
     \label{fig:Hier_Res}
\end{figure*}

The resulting $95\%$ CL exclusion bounds from CAST as well as the IAXO/IAXO+ projections are shown in \cref{fig:SummaryPlot,fig:quasiDegRes,fig:Hier_Res}. From \cref{eqn:exclusion-limits}, it follows that the shape of these bounds is entirely determined by the $\widetilde{N}_\mathrm{exp}$ prediction within the two-axion framework --- see Appendix~\ref{app:details-exclusion-limits} for a detailed explanation on its behavior.

\subsection{Two-axion discrimination}
Under the assumption of Asimov data, it is straightforward to obtain the minimal coupling $g_{a\gamma}^{\rm dis}$ for which a two-axion signal can be discriminated at $n_\sigma$ sigmas:
\begin{widetext}
\begin{equation}\label{eqn:g_dis}
g_{a\gamma}^{\rm dis} = \sqrt{n_\sigma}\left({2\int \dd\omega\, \widetilde{n}_\mathrm{exp}(\omega\,; m_2,\Delta m_{21}^2,\varphi) \log\left[\frac{\widetilde{P}(\omega\,;\, m_2,\Delta m_{21}^2,\varphi)}{h(m_2,\Delta m_{21}^2,\varphi) \widetilde{P}(\omega\,;\, \widehat{\widehat{m}}_2,0,0)}\right]}\right)^{-1/4}\,,
\end{equation}
\end{widetext}
where
\begin{equation}
h(m_2,\Delta m_{21}^2,\varphi) = \frac{\int \dd\omega\, \widetilde{n}_\mathrm{exp}(\omega\,; m_2,\Delta m_{21}^2,\varphi)}{\int \dd\omega\, \widetilde{n}_\mathrm{exp}(\omega\,; \widehat{\widehat{m}}_2,0,0)}\,.
\end{equation}
One can readily see that the function $h(m_2,\Delta m_{21}^2,\varphi)$ encodes the ratio between the total number of photon counts expected within the two-axion theory and the corresponding number within the single-axion theory that better fits the Asimov dataset. 

\subsubsection{Quasi-degenerate regime}

The results for the quasi-degenerate regime governed by the conversion probability in \cref{eqn:P-quasi-degenerate-general} are shown in the left panel of \cref{fig:quasiDegRes} for equal photon couplings $\varphi = \pi/4$ and $m_1 < m_{c,\,\mathrm{magnet}} \sim 10^{-2}$ eV. The light blue band marks the CAST-allowed region where IAXO would observe a statistically significant signal from two solar axions. However, within this band the two-axion spectrum  can mimic the single-axion one. The colored lines show $g_{a\gamma}^{\rm dis}$ at $3\sigma$ for different energy resolutions, and the shaded yellow region indicates where IAXO could discriminate the two-axion signal from the single-axion one within the CAST-allowed region. 

For small mass splittings, $\sqrt{\Delta m_{21}^2} \ll m_{c,\,\mathrm{Sun}}$, the cosine oscillation in the photon spectrum is too slow to produce a significant spectral distortion, and discrimination becomes increasingly difficult as $\sqrt{\Delta m_{21}^2} \to 0$. In this regime, the minimal coupling for which a
two-axion signal can be discriminated scales as
\begin{equation}\label{eqn:g_dis_scaling}
g_{a\gamma}^{\rm dis} \propto \frac{1}{\sin 2\varphi \cdot \Delta m_{21}^2}\,,
\end{equation}
corresponding to the parametric scaling indicated with a gray band in \cref{fig:quasiDegRes}. This follows from a simple analytical argument. Expanding the cosine in the probability of \cref{eqn:P-quasi-degenerate-general} for small $\Delta m_{21}^2 L_\mathrm{ES}/(2\omega)$, the deviation of the two-axion spectrum from the single-axion hypothesis scales as $s_{2\varphi}^2\,(\Delta m_{21}^2)^2$, so that $q_0 \propto g_{a\gamma}^4\,s_{2\varphi}^4\,(\Delta m_{21}^2)^4$, and setting $q_0 = n_\sigma^2$ yields \cref{eqn:g_dis_scaling}.

The optimal discrimination point (\reddiamond) in \cref{fig:quasiDegRes} corresponds to the value of $\sqrt{\Delta m_{21}^2}$ 
at which the combination of a large spectral distortion and a large number of photon counts gives rise to the best sensitivity in the coupling. Slightly smaller is the value of the mass splitting at which the oscillation minimum falls on the peak of the solar spectrum (\pinkcircle), corresponding to the weakest bound on the photon-axion coupling. Although the spectral difference between the single- and two-axion hypotheses is large there, the suppression of the total photon count also reduces the available statistics and limits the discrimination power. At the optimal discrimination point, one maximum of the cosine oscillation is aligned with the peak of the solar axion flux, see red curve in \cref{fig:appendix-spectra}.

\begin{figure}[ht]
     \includegraphics[width=1.02\columnwidth, trim={7pt 0 0 0}, clip]{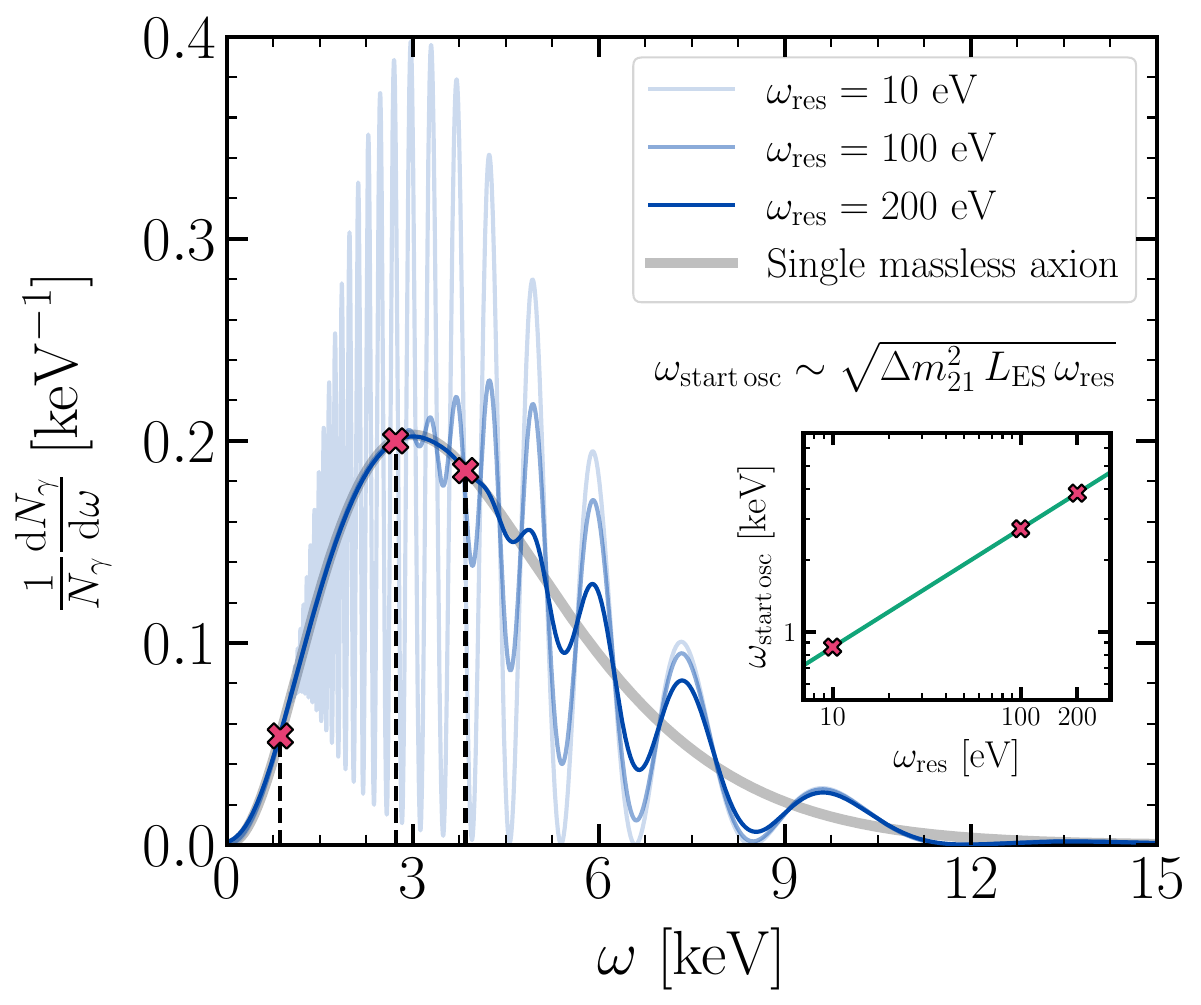}
     \caption{Effects of the finite energy resolution of the detector on the photon spectrum in the quasi-degenerate regime. The cross marks \scalebox{0.85}{\protect\pinkcross} indicate the energy that triggers the start of the oscillations for different resolutions. The inset in the bottom right further illustrates the dependence $\omega_\mathrm{start\ osc} \sim \sqrt{\omega_\mathrm{res}}$ for a fixed axion mass difference.\hfill\,}
     \label{fig:Resolution}
\end{figure}

For larger mass splittings, the oscillation of the spectrum becomes faster and thus the finite energy resolution $\omega_\mathrm{res}$ becomes the crucial experimental parameter. 
More precisely, the cosine phase $\Phi \equiv \Delta m_{21}^2 L_\mathrm{ES}/(2\omega)$ varies with energy as $|\delta\Phi| = \Delta m_{21}^2 L_\mathrm{ES}\,\delta\omega/(2\omega^2)$. When it completes more than a cycle across an energy interval $\delta\omega=\omega_\mathrm{res}$, \ie\ $|\delta\Phi| \sim 2\pi$, the detector averages over many oscillation cycles and the oscillatory pattern is washed out. 
This sets a critical energy $\omega_\mathrm{start\, osc} \sim \sqrt{\Delta m_{21}^2 L_\mathrm{ES}\,\omega_\mathrm{res}}$ above which the oscillations of the spectrum start to be resolved, as indicated with the cross mark \scalebox{0.85}{\pinkcross} in \cref{fig:Resolution} for different energy resolutions. It is thus clear that improving the energy resolution of the detector shifts $\omega_\mathrm{start\, osc}$ to lower energies, extending the range of $\sqrt{\Delta m_{21}^2}$ for which IAXO can discriminate two axions from one. Quantitatively, the maximum $\sqrt{\Delta m_{21}^2}$ for which IAXO can discriminate two axions within the CAST-allowed region scales as $\omega_\mathrm{res}^{-1/2}$, as indicated  in \cref{fig:quasiDegRes}  by an arrow pointing from the limit for $\omega_{\rm res}=200\;\hbox{eV}$ to the one for $\omega_{\rm res}=10\;\hbox{eV}$. The energy resolution of the X-ray detector is therefore the main experimental handle to extend the two-axion discrimination reach in the quasi-degenerate regime.

\begin{figure}[t]
     \includegraphics[width=1.0\columnwidth, trim={7pt 0 0 0}, clip]{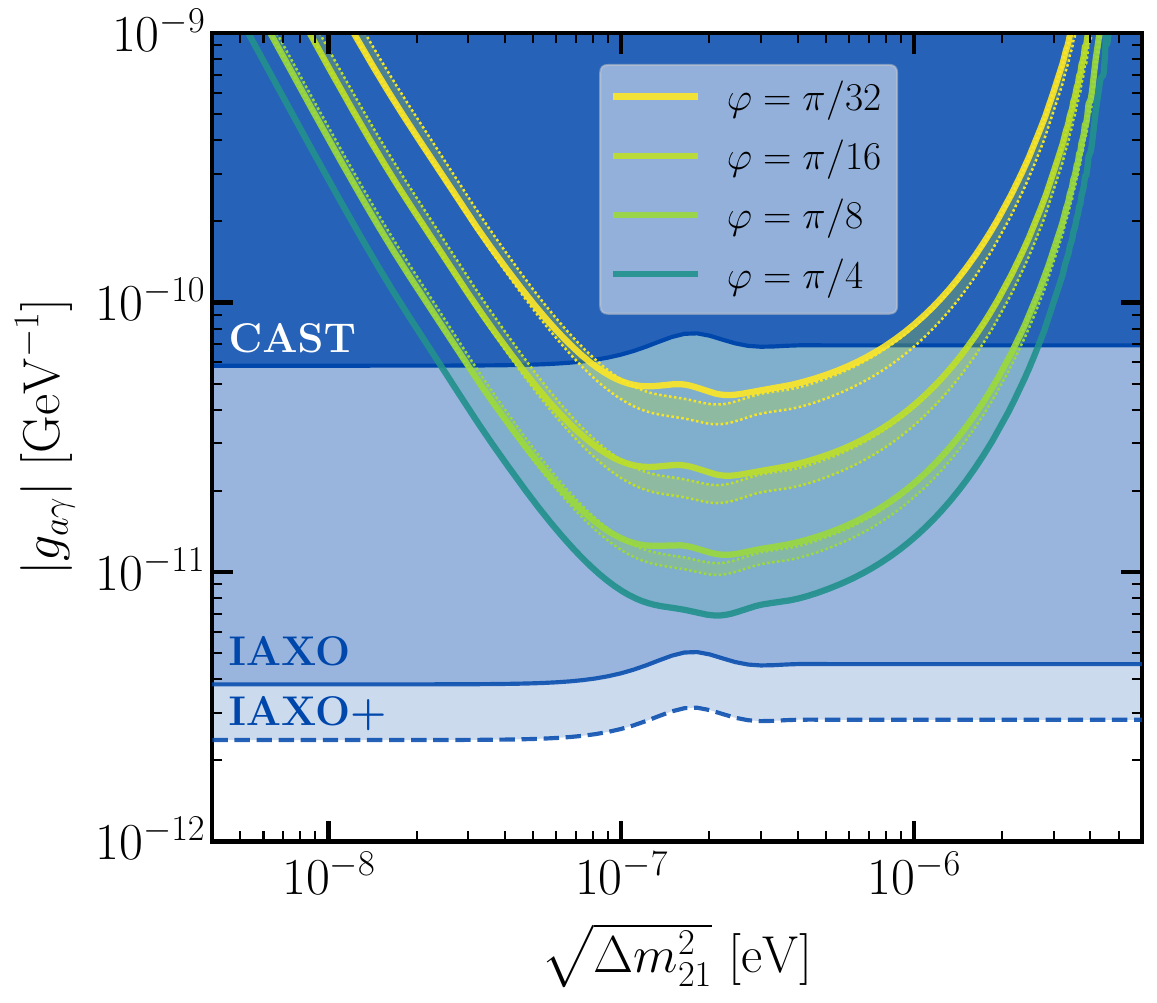}
     \caption{Dependence of the two-axion discrimination region with the mixing angle in the quasi-degenerate regime. The solid lines correspond to the full numerical results at each given angle $\varphi = \pi/4,\, \pi/8,\, \pi/16,\, \pi/32$. The lower dotted envelopes illustrate the scaling $g^{\rm dis}_{a\gamma} \propto 1/s_{2\varphi}$ for small mass differences, while the upper envelopes approximately follow $g_{a\gamma}^{\rm dis} \propto (2 - s_{2\varphi}^2)^{1/4}/s_{2\varphi}$ for large mass splittings.\hfill\,}
     \label{fig:quasi_Angles}
\end{figure}

The dependence on the mixing angle $\varphi$ is shown in \cref{fig:quasi_Angles}. Since discrimination relies on the cosine interference term in the conversion probability, which is proportional to the product of both eigenstates' photon couplings, $(c_\varphi\, s_\varphi)^2 \propto s_{2\varphi}^2$, the discrimination power peaks at $\varphi = \pi/4$ and vanishes as $\varphi \to 0$ or $\varphi \to \pi/2$, where one eigenstate decouples from photons. In the two limiting regimes the scaling takes analytic forms: at small $\sqrt{\Delta m_{21}^2}$, it follows immediately from \cref{eqn:g_dis_scaling} that $g_{a\gamma}^{\rm dis} \propto 1/s_{2\varphi}$ (lower dotted envelopes), while at large $\sqrt{\Delta m_{21}^2}$, where the cosine averages out, it becomes $g_{a\gamma}^{\rm dis} \propto (2 - s_{2\varphi}^2)^{1/4}/s_{2\varphi}$ (upper dotted envelopes). 

Finally, in \cref{fig:MassRegimes} we illustrate the discrimination region in the $\sqrt{\Delta m_{21}^2}$ \emph{vs.}~$m_1$ plane. In the quasi-degenerate mass regime, the discrimination region is centered around mass splittings of the order $m_{c,\,\mathrm{Sun}}$, precisely where the cosine oscillations become resolvable. The region is abruptly cut at $m_1 \sim 10^{-5}$~eV since astrophysical constraints already rule out such values of the axion mass in the relevant range of axion-photon couplings. Note that IAXO also can discriminate 2 axions in the intermediate window between the quasi-degenerate and hierarchical regimes where both axions have masses around $10^{-2}$~eV. For all the numerical analysis we always used the full conversion probability in \cref{eqn:Paxions2photon-vac}.
\subsubsection{Hierarchical regime}

The results for the hierarchical regime are shown in the right panel of \cref{fig:Hier_Res}, where $\sqrt{\Delta m_{21}^2} \simeq m_2$ since $m_1 \ll m_2$. The relevant conversion probability is \cref{eqn:P-hierarchical}, whose dominant feature is the $\sinc^2$ envelope from $m_2$, visible in the right panel of \cref{fig:spectra}.

The green line shows the minimum coupling at which IAXO can detect the axion mass from the $\sinc^2$ modulation of the spectrum, as obtained in Ref.~\cite{Dafni:2018tvj}. Above the green line, a massive axion signal is detectable and its mass measurable, but the number of axions contributing to the signal is undetermined. The yellow, orange, and red lines show $g_{a\gamma}^{\rm dis}$ at $3\sigma$-confidence level for energy resolutions $\omega_{\rm res} = 10,\,100,\,200$ eV, respectively. Above these curves, within the CAST-allowed region, IAXO could additionally establish the two-axion nature of the signal, discriminating it from a single massive axion.

Unlike in the quasi-degenerate regime and unlike the single-axion mass discovery reach, the energy resolution has almost no effect on the discrimination reach in the hierarchical regime. As visible in the right panel of \cref{fig:spectra}, the two-axion hierarchical spectrum differs from the single-axion one
through an overall shape distortion and a slight shift of the spectral peak, features that are not washed out by finite energy resolution.

Figure~\ref{fig:MassRegimes} shows a complementary view of the discrimination region. In the hierarchical mass regime, this is centered around mass splittings of the order $m_{c,\,\mathrm{magnet}}$, for which the sinc modulation in \cref{eqn:P-hierarchical} becomes relevant.

\subsubsection{Supernova axions}
Although IAXO was originally proposed to operate as a helioscope
\cite{IAXO:2019mpb}, its potential to detect axions produced in core-collapse supernovae has been recently investigated in Ref.~\cite{Carenza:2025uib}. SN axions could, in principle, allow us to probe different ranges of the axion mass splitting $\sqrt{\Delta m_{21}^2}$ since SNe are further away than the Sun, $L_{\rm SN}\gg L_{\rm ES}$, and the produced axions have larger energies, in the 10--250 MeV range. For SN axions with energies $\omega_\mathrm{peak\,SN} \simeq 40$ MeV, the critical mass splitting at which the cosine oscillation length becomes comparable to $L_\mathrm{SN} \sim \mathcal{O}(100\text{ pc})$
is $m_{c,\,\mathrm{SN}} \equiv \sqrt{\omega_\mathrm{peak\,SN}/L_\mathrm{SN}} \sim 10^{-9}$~eV, roughly two orders of magnitude smaller than $m_{c,\,\mathrm{Sun}}$. On the other hand, now the sinc modulation becomes relevant for axion masses of the order $m_{c,\,\mathrm{magnet\, SN} } \equiv \sqrt{\omega_\mathrm{peak\,SN}/L} \sim 1$~eV. 

The extreme temperature and density conditions in the SN core enhance the production of axions through their coupling to nucleons $g_{aN}$, with $N = p,n$ for protons and neutrons, respectively. The main production mechanisms \cite{Carenza:2023lci} are $NN$ bremsstrahlung $N+N \to N+N+a$, and pion-nucleon scattering $\pi+N \to a+N$. Taking into account these two processes, fitting expressions for the axion emission flux are provided in Ref.~\cite{Lella:2024hfk} and summarized in Appendix~\ref{app:SN-axions}. The poorly known relative contribution from pion-nucleon scattering is accounted for by a parameter, $\delta$, that encodes the uncertain pion abundance in the inner core regions. 
Accordingly, we consider three benchmark scenarios for  $\{\delta,\ g_{an},\ g_{ap}\}$: (i) $\{1,\ 1.3\times10^{-9},\ 0\}$, (ii) $\{0,\ 1.3\times10^{-9},\ 0\}$, and (iii) $\{1,\ 0,\ 6\times10^{-10}\}$, where $\delta=1$ corresponds to $Y_\pi \sim \mathcal{O}(1\%)$. In (i) and (ii) the neutron coupling saturates the neutron-star cooling bound~\cite{Buschmann:2021juv}, and they differ on the pion contribution. Scenario (iii) saturates the stronger SN1987A proton coupling bound~\cite{Lella:2023bfb}, probing the impact of a weaker coupling.

On the experimental side, we assume a gamma-ray detector efficiency $\varepsilon = 0.95$, following the prescription of Ref.~\cite{Carenza:2025uib}, and energy resolution $\omega_\mathrm{res} = 1$ MeV. 

In \cref{fig:SummaryPlot}, we have included (in pink) the two-axion discrimination region using input parameters   from the SN candidate Spica/$\alpha$ Virginis, located at $L_\mathrm{SN} = 0.077$ kpc \cite{Mukhopadhyay:2020ubs} from Earth. As anticipated, the reach of the discrimination region at IAXO is extended to lower values of the mass splitting $\sqrt{\Delta m_{21}^2}$.

Although with detection of SN axions one can in principle access different mass splittings than solar axions and hence could offer a complementary probe, in practice discrimination of the two-axion scenario would be extremely difficult at IAXO. 
Figure~\ref{fig:SummaryPlot} shows the discrimination region allowed by CAST for the benchmark scenario (i). This region is compared with the other two benchmarks in right-hand panel of \cref{fig:SN_discovery} (Appendix~\ref{app:SN-axions}). For benchmark (ii) the region shrinks, since setting $\delta = 0$ removes the pion-nucleon contribution and reduces the flux at higher energies, while for (iii) the smaller axion-nucleon coupling lowers the total flux so much that discrimination would require a value of $g_{a\gamma}$ already excluded by CAST.

Additionally, the study of SN axions comes with several difficulties. First of all, SN explosions occur very rarely (roughly one per century) and, secondly, the axion flux on Earth lasts only for a few seconds. On that regard, the observation of pre-SN neutrinos and the subsequent fast response of the IAXO tracking system to point at a specific target would be essential. From the theoretical point of view, a better understanding of the pion scattering relative contribution and the SN axion spectrum is mandatory in order to identify the regions of interest in the parameter space.

\section{Generalization to \texorpdfstring{$N$}{N} axions} 
\label{sec:generalization_to_n_axions}

The analysis presented in the previous sections focuses on the two-axion system, which, as we argue below, often captures the essential phenomenology of the general $N$-axion scenario for practical helioscope purposes. Consider $N$ axion mass eigenstates $\{a_1,\ldots,a_N\}$ with masses $m_1 < m_2 < \cdots < m_N$. The axion--photon interaction Lagrangian reads
\begin{equation}\label{eq:Lag-N-axions}
    \Lag_{a\gamma} = -\frac{g_{a\gamma}}{4}\left(\sum_{n=1}^N c_n\, a_n\right) F_{\mu\nu}\widetilde{F}^{\mu\nu}
    \equiv -\frac{g_{a\gamma}}{4}\,\ag\,F_{\mu\nu}\widetilde{F}^{\mu\nu}\,,
\end{equation}
where $g_{a\gamma} \equiv \sqrt{\sum_{i=1}^N g_i^2}$ is the total effective axion-photon coupling, $\ag \equiv \sum_{n} c_n \, a_n$ is the interaction state, and $c_n$ are the normalized couplings of each mass eigenstate to photons, satisfying $\sum_{n=1}^N c_n^2 = 1$. This generalizes the two-axion mixing angle where $c_1 = c_\varphi$, $c_2 = s_\varphi$ and $2c_1^2c_2^2 = (1/2) s_{2\varphi}^2$.
The corresponding conversion probability in vacuum then takes the form
\begin{widetext}
\begin{equation}\label{eq:P-N-axions}
    P_{\ag\to\gamma}^{(N)}
    = \left(\frac{g_{a\gamma} B L}{2}\right)^{\!2}
    \left[
        \sum_{i=1}^N c_i^4\,\sinc^2\!\left(\frac{m_i^2 L}{4\omega}\right)
        + 2\!\sum_{i<j}^N c_i^2\,c_j^2\,
        \sinc\!\left(\frac{m_i^2 L}{4\omega}\right)\sinc\!\left(\frac{m_j^2 L}{4\omega}\right)
        \cos\!\left(\frac{\Delta m_{ji}^2\, L_\mathrm{ES}}{2\omega}\right)
    \right]\,,
\end{equation}
\end{widetext}
with $\Delta m_{ji}^2 \equiv m_j^2 - m_i^2 > 0$ for $i < j$. This expression contains $N$ diagonal terms and $\binom{N}{2}$ pairwise interference terms, and reduces to \cref{eqn:Paxions2photon-vac} for $N=2$.

Although \cref{eq:P-N-axions} is the complete conversion probability, in most realistic mass configurations it simplifies considerably. Whether each term produces a detectable spectral signature is governed by two mass scales. The \emph{sinc window}, set by the magnet length $L$, controls the diagonal terms: eigenstates with $m_i \ll m_{c,\,\mathrm{magnet}}$ contribute with $\sinc \simeq 1$ (mass independent), those with $m_i \sim m_{c,\,\mathrm{magnet}}$ produce a measurable sinc modulation, and those with $m_i \gg m_{c,\,\mathrm{magnet}}$ decouple entirely ($\sinc \to 0$). The \emph{cosine window}, set by $L_\mathrm{ES}$ and the energy resolution $\omega_\mathrm{res}$, controls the interference terms: pairs with $\sqrt{\smash[b]{\Delta m_{ij}^2}} \ll m_{c,\,\mathrm{Sun}}$ have $\cos \simeq 1$ and are spectrally indistinguishable from a single axion, pairs with $\sqrt{\smash[b]{\Delta m_{ij}^2}} \gg m_{c,\,\mathrm{Sun}}$ are washed out by finite resolution ($\langle\cos\rangle \to 0$), and only pairs in the intermediate window $\sqrt{\smash[b]{\Delta m_{ij}^2}} \sim m_{c,\,\mathrm{Sun}}$ produce a resolvable oscillation.

Two reductions simplify \cref{eq:P-N-axions} before considering specific mass patterns. First, any eigenstate with $m_i \gg m_{c,\,\mathrm{magnet}}$ has $\sinc \to 0$ and decouples completely from the signal. Second, any subset of axions with mutual mass differences $\sqrt{\smash[b]{\Delta m_{ij}^2}} \ll m_{c,\,\mathrm{Sun}}$ is spectrally indistinguishable for IAXO, and thus effectively degenerate. 
The interference terms among them have $\cos \simeq 1$, so their contributions combine into a perfect square $\left(\sum c_j^2\right)^2 \equiv C_\alpha^4$ (sum over axions in the group), and the entire subset is merged into an effective axion $\alpha$ with coupling $C_\alpha$.
Physically, this follows from those subsets being degenerate in practice for IAXO. The only relevant combination within the degenerate subset is the one that couples to photons, ${C_\alpha A_\alpha\equiv\,\sum_j c_j\,a_j}$ (a good approximate mass eigenstate), while all other combinations are decoupled.
 After these reductions, $\tilde{N}$ effective groups remain (labeled $\alpha = 1,\ldots,\tilde{N}$ in order of increasing mass) with $\sum_\alpha C_\alpha^2 \leq 1$ (equality when no eigenstate has $m_i \gg m_{c,\,\mathrm{magnet}}$), and the spectral signal is governed by the $\tilde{N}$ diagonal sinc terms and $\binom{\tilde{N}}{2}$ cosine terms among groups. Two prominent limiting cases each reduce to the two-axion formulae of \cref{sec:results}.
 
\emph{Quasi-degenerate class.} All $\tilde{N}$ groups lie well below $m_{c,\,\mathrm{magnet}}$ so $\sinc \simeq 1$ for all.\footnote{More generally, if some groups have $m_\alpha \sim m_{c,\,\mathrm{magnet}}$, the formula holds with $\mathcal{D} \to \sum_\alpha C_\alpha^4\,\sinc_\alpha^2$ and $2C_A^2 C_B^2\cos \to 2C_A^2 C_B^2\,\sinc_A\,\sinc_B\,\cos$. The mapping to \cref{eqn:P-quasi-degenerate-general} then also holds with these modified values.}
 Exactly one pair between groups $(A, B)$ has $\sqrt{\smash[b]{\Delta m_{AB}^2}} \sim m_{c,\,\mathrm{Sun}}$, while all remaining $\binom{\tilde{N}}{2} - 1$ pairs have larger splittings and  wash out ($\langle \cos \rangle \to 0$). Defining $\mathcal{D}^4 \equiv \sum_\alpha C_\alpha^4$, the probability reduces to
\begin{equation}\label{eq:P-N-QD}
    P_{\ag\to\gamma}^{(N)} \simeq \left(\frac{g_{a\gamma} B L}{2}\right)^{\!2}
    \Bigg[\mathcal{D}^4 + 2\,C_A^2\,C_B^2\,\cos\!\left(\frac{\Delta m_{AB}^2\, L_\mathrm{ES}}{2\omega}\right)\Bigg]\,.
\end{equation}
This maps exactly to the $N=2$ quasi-degenerate formula in \cref{eqn:P-quasi-degenerate-general} with $g_{\rm eff}^2 = g_{a\gamma}^2(\mathcal{D}^4 + 2C_A^2 C_B^2)$ and mixing angle
\begin{equation}\label{eq:sin2phi-eff}
    \sin^2 2\varphi_{\rm eff} = \frac{4\,C_A^2\,C_B^2}{\mathcal{D}^4 + 2\,C_A^2\,C_B^2}\,.
\end{equation}
The discrimination analysis of \cref{sec:results} therefore applies with these effective parameters.

\emph{Hierarchical class.} All $\binom{\tilde{N}}{2}$ splittings between groups satisfy $\sqrt{\smash[b]{\Delta m^2}} \gg m_{c,\,\mathrm{Sun}}$, so all cosine terms wash out ($\langle \cos \rangle \to 0$). The heaviest group has $m_{\tilde{N}} \sim m_{c,\,\mathrm{magnet}}$, while all others satisfy $m_\alpha \ll m_{c,\,\mathrm{magnet}}$ ($\sinc \simeq 1$). Defining $\mathcal{D}_L^4 \equiv \sum_{\alpha < \tilde{N}} C_\alpha^4$, the probability reduces to
\begin{equation}\label{eq:P-N-hier}
    P_{\ag\to\gamma}^{(N)} \simeq \left(\frac{g_{a\gamma} B L}{2}\right)^{\!2}
    \Bigg[\mathcal{D}_L^4 + C_{\tilde{N}}^4\,\sinc^2\!\left(\frac{m_{\tilde{N}}^2\, L}{4\omega}\right)\Bigg]\,.
\end{equation}
This maps exactly to the $N=2$ hierarchical formula in \cref{eqn:P-hierarchical} with $g_{\rm eff}^2 = g_{a\gamma}^2(\mathcal{D}_L^2 + C_{\tilde{N}}^2)^2$ and $s_{\varphi,\rm eff}^2 = C_{\tilde{N}}^2/(\mathcal{D}_L^2 + C_{\tilde{N}}^2)$. The discrimination analysis of \cref{sec:results} therefore applies directly.

If neither condition holds, \ie\ no splitting between groups falls within the cosine window and no group lies near $m_{c,\,\mathrm{magnet}}$, all spectral features are absent and the signal is indistinguishable from a single axion. If several mass splittings are of the order $m_{\rm c,\, Sun}$, or several masses are of the order of $m_{\rm c,\,magnet}$, the $N$-axion system cannot be reduced to the 2-axion one and a dedicated analysis is needed.
\begin{figure}[ht]
     \includegraphics[width=1.0\columnwidth, trim={7pt 0 0 0}, clip]{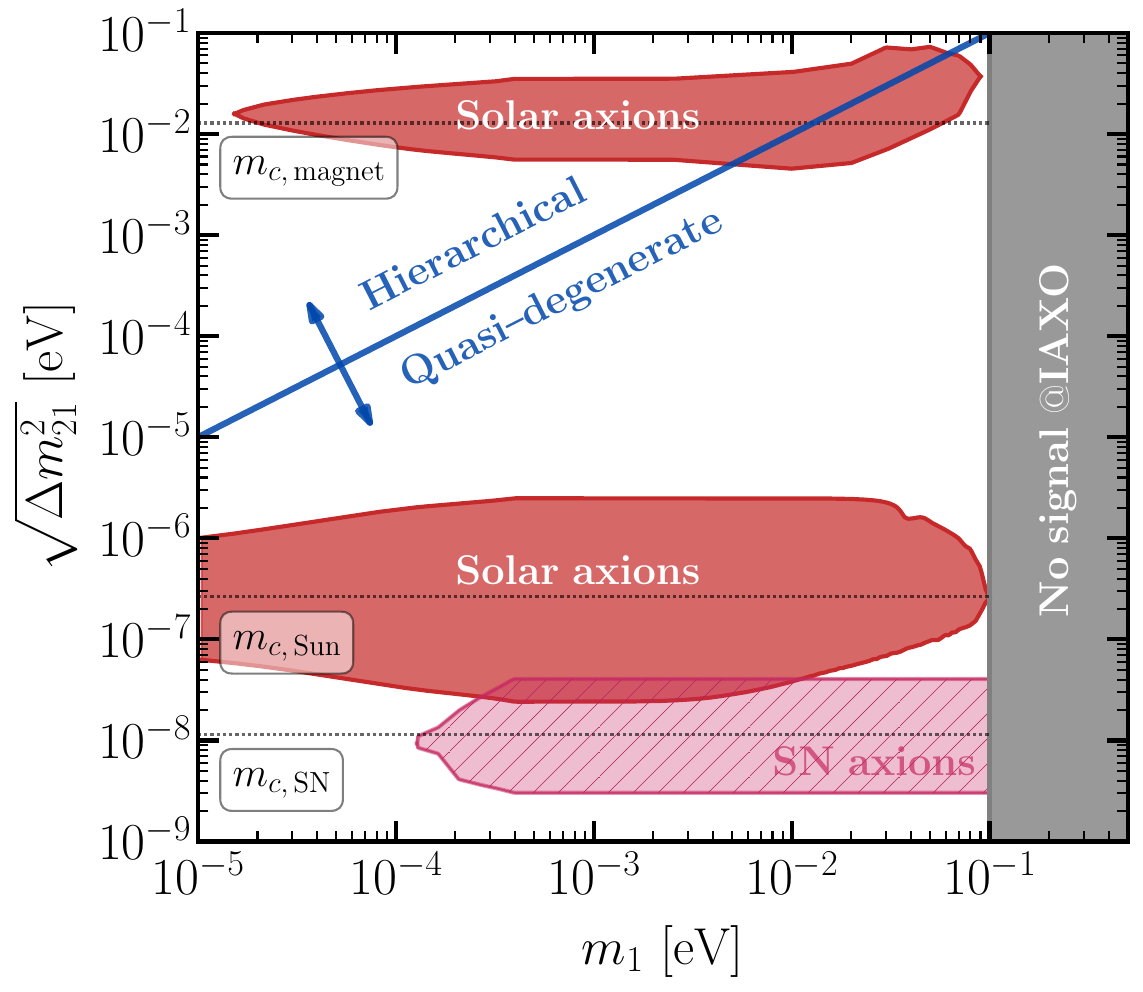}
     \caption{Two-axion discrimination regions in the $\sqrt{\Delta m_{21}^2}$ \emph{vs.} $m_1$ plane, with $m_1$ the mass of the lightest axion, obtained from the analysis with solar axions (red) and SN axions (hatched pink). Each of them is centered around a certain critical mass splitting depending on which oscillatory term is dominant in \cref{eqn:Paxions2photon-vac}. The blue line indicates the separation between the quasi-degenerate regime $\sqrt{\Delta m_{21}^2} \simeq m_{1,2}$ and the hierarchical regime $m_1 \ll \sqrt{\Delta m_{21}^2} \simeq m_2$. The region in gray points out the mass interval where IAXO will not be able to detect any signal. \hfill\,}
     \label{fig:MassRegimes}
\end{figure}
It is interesting to explore how the photon signal scales with $N$ when all axions couple similarly, $c_n \sim \mathcal{O}(1/\sqrt{N})$, so that $\sum_n c_n^2 = 1$ is satisfied for any $N$. For masses well below $m_{c,\,\mathrm{magnet}}$, \cref{eq:P-N-axions} gives
\begin{equation}\label{eq:scaling-N}
    P_{\ag\to\gamma}^{(N)} \simeq \begin{cases}
        P_{\ag\to\gamma}^{\mathrm{1\text{-}axion}} &\text{Coherent}\quad\,\,\,\, \sqrt{\smash[b]{\Delta m_{ij}^2}} \ll m_{c,\,\mathrm{Sun}}\,,\\[4pt]
        \dfrac{1}{N}\,P_{\ag\to\gamma}^{\mathrm{1\text{-}axion}} &\text{Incoherent}\quad \sqrt{\smash[b]{\Delta m_{ij}^2}} \gg m_{c,\,\mathrm{Sun}}\,,
    \end{cases}
\end{equation}
where $P_{\ag\to\gamma}^{\mathrm{1\text{-}axion}} = (g_{a\gamma}BL/2)^2$. The $N$-axion signal is therefore always bounded from above by the single-axion result at the same $g_{a\gamma}$, with equality in the coherent regime and a $1/N$ suppression\footnote{References~\cite{Chadha-Day:2021uyt,Chadha-Day:2023wub,deGiorgi:2025ldc} adopt a different normalization, $c_n \sim \mathcal{O}(1)$. This yields apparent $N^2$ and $N$ enhancements relative to a single axion at their $g_{a\gamma}$ (which is $\sqrt{N}$ times larger than ours). Our convention $\sum_n c_n^2 = 1$ sets $g_{a\gamma}$ as the coupling of the canonically normalized interaction state $a_\gamma$ and has the correct single-axion limit: when all $N$ axions are degenerate, the mass matrix is proportional to the identity and one can always rotate to a basis where $N-1$ combinations decouple completely, leaving a single coupled state $a_\gamma$ with coupling $g_{a\gamma}$.} once all mass splittings exceed $m_{c,\,\mathrm{Sun}}$ and the photons converted from each eigenstate interfere incoherently.

\section{Conclusions} 
\label{sec:discussion_and_outlook}

The existence of multiple axion species is a generic prediction of well-motivated extensions of the SM, from string compactifications to multi-axion solutions to the strong CP problem. In this work, we have addressed a natural question raised by this theoretical landscape: if a next-generation helioscope detected a signal, would we be able to disentangle whether it originates from one or multiple axions? We have shown that IAXO will indeed be able to tell them apart in some regions of the parameter space.

Axions produced in the Sun can convert into photons within the transverse magnetic field of the helioscope, in such a way that the detected X-ray spectrum carries enough information to discriminate a two-axion theory from the single-axion hypothesis in two different mass regimes. 
In the quasi-degenerate regime, $\Delta m_{21}^2 \ll m_{1,2}^2$, discrimination is driven by the cosine modulation of the spectrum resulting from \emph{axion flavor oscillations} along the Earth-Sun distance, and is most pronounced for mass splittings $\sqrt{\Delta m_{21}^2} \sim m_{c,\,\mathrm{Sun}} \sim 10^{-7}$ eV, the scale at which the oscillation length becomes comparable to $L_\mathrm{ES}$. In the hierarchical regime $m_1^2 \ll m_2^2 \simeq \Delta m_{21}^2$, the relevant mass scale is instead $m_{c,\,\mathrm{magnet}} \sim 10^{-2}$ eV, for which the $\sinc^2$ modulation, controlled by the magnet length, becomes resolvable and shapes the photon spectrum. The ability of future helioscopes to distinguish the two-axion from the single-axion hypothesis across two distinct mass-splitting regimes further motivates the physics case of BabyIAXO~\cite{IAXO:2020wwp} and IAXO/IAXO+~\cite{IAXO:2025ltd}.

From the experimental point of view, an important result of our study is that the energy resolution of the X-ray detectors $\omega_\mathrm{res}$ is the key experimental handle for extending the two-axion discrimination reach to larger mass splittings in the quasi-degenerate regime. 
The maximum mass splitting accessible to IAXO depends on the detector capability to resolve the cosine oscillations as they become faster, and scales parametrically as $\sqrt{\Delta m_{21}^2}\big|_\mathrm{max} \propto \omega_\mathrm{res}^{-1/2}$. In contrast, in the hierarchical regime the distinction between the two-axion and the single-axion spectra relies on an overall distortion of the spectral shape rather than on rapid oscillations, and thus the resolution of the detector has almost no effect in the discrimination reach. 
The same conclusion holds for the ability of IAXO to measure the single axion mass, where improving energy resolution also yields marginal gains, since the $\sinc^2$ modulation at high masses not only generates fast oscillations but also suppresses their amplitude.

In addition, we have also explored whether supernova axions could provide a complementary probe of the two-axion parameter space, exploiting the larger Earth-SN distances and the higher axion energies. Although photons from SN axions are extremely rare and present formidable experimental challenges, they would exhibit an oscillatory pattern for a wide window of mass splittings of $\sqrt{\Delta m_{21}^2} \sim 10^{-9}$--$10^{-7}$~eV. In a more conservative scenario the discrimination region shrinks or is even excluded by CAST. Unlike the solar case, where the detector energy resolution sets the reach, the limiting factor for SN axions is not instrumental but astrophysical: our large uncertainty on the initial SN axion flux, encoded in the parameter $\delta$. Developments in supernova physics that pin down $\delta$ would therefore help clarify the ability of IAXO to disentangle one from two axions with SN data.

A natural question is whether the two-axion systems in our discrimination regions could solve the strong CP problem, which requires checking the QCD axion sum rule~\cite{Gavela:2023tzu}. In practice, helioscope data would fix only some of the parameters of the axion system (the mass splitting and photon couplings), but a full test of the sum rule would require additional input, such as the individual $E/N$ of each axion or, equivalently, the gluonic couplings. Experiments such as CASPEr-Electric~\cite{Budker:2013hfa} could provide this additional input by measuring the individual axion masses and gluonic couplings.
In the quasi-degenerate regime, the discrimination regions are compatible with both axions lying within the standard QCD band. 
In the hierarchical regime, discrimination requires the heaviest axion to have $E/N \gtrsim 7$ (see overlap in \cref{fig:SummaryPlot}), or to come from more exotic photophilic constructions such as clockwork models~\cite{Giudice:2016yja,Farina:2016tgd} or the light $\mathcal{Z_N}$ axion~\cite{Hook:2018jle,DiLuzio:2021pxd,DiLuzio:2021gos}.

Finally, although our analysis has focused on the two-axion system, it is important to emphasize that the relevant phenomenology at IAXO of a much larger class of $N$-axion patterns (with $N>2$) is already captured by our results. As shown in \cref{sec:generalization_to_n_axions}, an $N$-axion system reduces to an effective two-axion problem whenever some of the axion masses lie close to one of the two critical scales of the analysis, $m_{c,\,\mathrm{magnet}}$ or $m_{c,\,\mathrm{Sun}}$. This makes our discrimination reach directly applicable to a wide variety of motivated multi-axion patterns arising in string theory or extra-dimensional models, among others.

Beyond establishing the multi-axion nature of a signal, the helioscope measurement of $\Delta m_{21}^2$ would also provide a first handle on the mass scale of the underlying axion system. This is complementary to resonant haloscopes, such as ADMX~\cite{ADMX:2025vom}, which rely on a mass resonance and are thus sensitive to each axion mass eigenstate individually. While the helioscope fixes the mass splitting, a haloscope detection of one eigenstate would set its absolute mass, so that $\Delta m_{21}^2$ would provide a definite target for the second axion. A direct search at that mass would then test whether the second axion also makes up part of the dark matter. The combination of helioscope and haloscope information is therefore a promising avenue to unravel the multi-axion landscape that may lie beyond the SM.

\appendix
 {\bf Acknowledgments.}---%
We thank Axel Lindner, Javier Redondo, Igor Irastorza, Itay Bloch, Ciaran O'Hare, and Xingzhou Yu for helpful discussions. 
We specially appreciate the open-science practices of the authors of Ref.~\cite{Dafni:2018tvj}.
P.Q. acknowledges support by the European Union's Horizon 2020 research and innovation programme under the Marie Sk\l odowska-Curie Postdoctoral Fellowship grant agreement No 101207780 - AxionCount and  No 860881-HIDDeN, and under 
the Marie Sklodowska-Curie Staff Exchange grant agreement No 101086085-ASYMMETRY (B.G. and P.Q.). 
The work of B.G. is
supported by the U.S. Department of Energy under grant
number DE-SC0009919. Some of the work was carried out while B.G. was a Visiting Professor at Universit\`a di Roma, La Sapienza, and a Visiting Scientist at the Instituto de F\'isica Te\'orica, Madrid, and he is grateful for their hospitality. 
C.M. specially thanks UC San Diego and CERN Theoretical Physics Department, where part of this work was carried out.

\bibliographystyle{utphys}
\bibliography{Countbiblio}

\newpage 
\onecolumngrid

\appendix
\onecolumngrid
\section{Hypothesis testing}\label{app:hypothesis-testing}
In this appendix, we provide further details on the statistical methodology applied to derive the main results of this work. In Appendix \ref{subapp:Binned-analysis}, we first introduce the likelihood ratio test statistic $q_0$ in the context of a binned analysis. Then, it will be straightforward to take the small bin size limit in Appendix \ref{subapp:Unbinned-analysis} in order to reproduce the formulae previously presented in the main text.

\subsection{Binned likelihood analysis}\label{subapp:Binned-analysis}
The binned likelihood function consists of the product of the Poisson probability distribution, $\mathscr{P}(k\,|\lambda) = \frac{\lambda^k}{k!}e^{-\lambda}$, for the number of observed photons in each energy bin $i$, $N_\mathrm{obs}^i$, given an expected number $N_\mathrm{exp}^i(g_{a\gamma},m_2,\Delta m_{21}^2,\varphi)$:
\begin{equation}
\mathcal{L}(g_{a\gamma},m_2,\Delta m_{21}^2,\varphi) = \prod_{i=1}^{N_\mathrm{bins}} \mathscr{P}\left[N_\mathrm{obs}^i\,|\, N_\mathrm{exp}^i(g_{a\gamma},m_2,\Delta m_{21}^2,\varphi)\right]\,.
\end{equation}
It is straightforward to check that, in the large sample limit, the log-likelihood function
simply follows a $\chi^2$ distribution, 
\begin{equation}
-2 \log\mathcal{L}(g_{a\gamma},m_2,\Delta m_{21}^2,\varphi) = \sum_{i=1}^{N_\mathrm{bins}}\frac{\left[N_\mathrm{obs}^i-N_\mathrm{exp}^i(g_{a\gamma},m_2,\Delta m_{21}^2,\varphi)\right]^2}{N_\mathrm{exp}^i(g_{a\gamma},m_2,\Delta m_{21}^2,\varphi)}\,.
\end{equation}
The expected number of photon counts in bin $i$ within the two-axion theory is given by
\begin{equation}\label{eqn:Nexp-bin-i}
N_\mathrm{exp}^i(g_{a\gamma},m_2,\Delta m_{21}^2,\varphi) = g_{a\gamma}^4 \int_{\omega_i}^{\omega_{i+1}} \dd \omega\, C(\omega)\, \frac{\dd \widetilde{\Phi}}{\dd \omega}(\omega)\, \widetilde{P}(\omega\,; m_2,\Delta m_{21}^2,\varphi)\,,
\end{equation}
with $\frac{\dd \widetilde{\Phi}}{\dd \omega}(\omega)$ and $\widetilde{P}(\omega\,; m_2,\Delta m_{21}^2,\varphi)$ defined in \cref{eqn:solar-flux-tilde,eqn:probability-tilde}, respectively.
Analogously, we have
\begin{equation}
N_\mathrm{exp}^i(g_{a\gamma},m_2,\Delta m_{21}^2,\varphi) \equiv g_{a\gamma}^4\; \widetilde{N}_\mathrm{exp}^i(m_2,\Delta m_{21}^2,\varphi)\,.
\end{equation}

As stated before, the null hypothesis $H_0$ and the alternative hypothesis $H_1$ correspond to
\begin{align}
&H_0: \qquad \Delta m_{21}^2 = 0 \quad \,\,\text{ or }\,\, \quad \varphi=0,\pi/2 \;\;\implies\;\; \text{1-axion theory}\,, \\[5pt]
&H_1: \qquad \Delta m_{21}^2 \neq 0 \quad \text{ and } \quad \varphi\neq 0,\pi/2 \;\;\implies\;\; \text{2-axion theory}\,.
\end{align}
In order to test the single-axion hypothesis against the two-axion theory, we define the likelihood ratio test statistic $q_0$ as \cite{Cowan:2010js} 
\begin{equation}
q_0 = \begin{cases} \;-2\log \Lambda\,, \;\;\quad \Delta \widehat{m}_{21}^2 > 0\\[5pt] \;0\,, \qquad\quad\;\;\;\;\;\; \Delta \widehat{m}_{21}^2 < 0 \end{cases}, \qquad\text{with}\quad \Lambda = \frac{\mathcal{L}(\widehat{\widehat{g}}_{a\gamma},\widehat{\widehat{m}}_2,0,0)}{\mathcal{L}(\widehat{g}_{a\gamma},\widehat{m}_2,\Delta \widehat{m}_{21}^2,\widehat{\varphi})}\,.
\end{equation}
The parameters $(\widehat{\widehat{g}}_{a\gamma},\widehat{\widehat{m}}_2)$ in the numerator are the conditional MLE for the specified $\Delta m_{21}^2 = 0$ and $\varphi = 0$, while $(\widehat{g}_{a\gamma},\widehat{m}_2,\Delta \widehat{m}_{21}^2,\widehat{\varphi})$ denote the (unconditional) MLE. From the maximization condition on the coupling, it is straightforward to check that
\begin{align}
\sum_{i=1}^{N_\mathrm{bins}} N_\mathrm{obs}^i = \sum_{i=1}^{N_\mathrm{bins}} N_\mathrm{exp}^i (\widehat{\widehat{g}}_{a\gamma},\widehat{\widehat{m}}_2,0,0) &\;\;\implies\;\; \widehat{\widehat{g}}_{a\gamma}^4 = \frac{\sum_{i=1}^{N_\mathrm{bins}} N_\mathrm{obs}^i}{\sum_{i=1}^{N_\mathrm{bins}} \widetilde{N}_\mathrm{exp}^i (\widehat{\widehat{m}}_2,0,0)}\,,\\[5pt]
\sum_{i=1}^{N_\mathrm{bins}} N_\mathrm{obs}^i = \sum_{i=1}^{N_\mathrm{bins}} N_\mathrm{exp}^i (\widehat{g}_{a\gamma},\widehat{m}_2,\Delta\widehat{m}_{21}^2,\widehat{\varphi}) &\;\;\implies\;\; \widehat{g}_{a\gamma}^4 = \frac{\sum_{i=1}^{N_\mathrm{bins}} N_\mathrm{obs}^i}{\sum_{i=1}^{N_\mathrm{bins}} \widetilde{N}_\mathrm{exp}^i (\widehat{m}_2,\Delta\widehat{m}_{21}^2,\widehat{\varphi})}\,,
\end{align}
and thus
\begin{equation}\label{eqn:ratio-MLE-g}
\sum_{i=1}^{N_\mathrm{bins}} N_\mathrm{exp}^i (\widehat{\widehat{g}}_{a\gamma},\widehat{\widehat{m}}_2,0,0) = \sum_{i=1}^{N_\mathrm{bins}} N_\mathrm{exp}^i (\widehat{g}_{a\gamma},\widehat{m}_2,\Delta\widehat{m}_{21}^2,\widehat{\varphi}) \;\;\implies\;\; \frac{\widehat{\widehat{g}}_{a\gamma}^4}{\widehat{g}_{a\gamma}^4} = \frac{\sum_{i=1}^{N_\mathrm{bins}} \widetilde{N}_\mathrm{exp}^i (\widehat{m}_2,\Delta\widehat{m}_{21}^2,\widehat{\varphi})}{\sum_{i=1}^{N_\mathrm{bins}} \widetilde{N}_\mathrm{exp}^i (\widehat{\widehat{m}}_2,0,0)}\,.
\end{equation}
At this point, we can make use of the Asimov dataset \cite{Cowan:2010js}. The latter is generated under the assumption of the alternative hypothesis $H_1$ (the two-axion theory). This means that the observed data in a future experiment such as IAXO is assumed to be \emph{exactly} described by the alternative hypothesis, that is, there are no (Poisson) fluctuations in the dataset. 
Therefore, the observed number of events in bin $i$ coincides with the number of events $N_\mathrm{exp}^i$ predicted by the two-axion theory, and the MLE correspond to the true values of the parameters:
\begin{equation}\label{eqn:Asimov-data-cond}
N_\mathrm{obs}^i = N_\mathrm{exp}^i(g_{a\gamma},m_2,\Delta m_{21}^2,\varphi) = N_\mathrm{exp}^i(\widehat{g}_{a\gamma},\widehat{m}_2,\Delta \widehat{m}_{21}^2,\widehat{\varphi})\,.
\end{equation}
The key point is that, under the assumption of the Asimov dataset, we can quantify the median discrimination significance of the two-axion theory against the single-axion hypothesis simply as $n_\sigma^2 = q_0$, with $q_0$ depending on the parameters of the theory. From this condition, it is then straightforward to compute the minimal coupling $g_{a\gamma}^\mathrm{dis}$ for which the two-axion signal can be discriminated at $n_\sigma$ sigmas as a function of the remaining model parameters
\begin{equation}\label{eqn:gdis-binned-analysis}
g_{a\gamma}^\mathrm{dis} = \sqrt{n_\sigma}\left(2\sum_{i=1}^{N_\mathrm{bins}} \widetilde{N}_\mathrm{exp}^i(m_2,\Delta m_{21}^2,\varphi) \log\left[\frac{\widetilde{N}_\mathrm{exp}^i(m_2,\Delta m_{21}^2,\varphi)}{h(m_2,\Delta m_{21}^2,\varphi)\widetilde{N}_\mathrm{exp}^i(\widehat{\widehat{m}}_2,0,0)}\right]\right)^{-1/4}\,,
\end{equation}
where
\begin{equation}
h(m_2,\Delta m_{21}^2,\varphi) \equiv \frac{\widehat{\widehat{g}}_{a\gamma}^4}{g_{a\gamma}^4} = \frac{\sum_{i=1}^{N_\mathrm{bins}} \widetilde{N}_\mathrm{exp}^i (m_2,\Delta m_{21}^2,\varphi)}{\sum_{i=1}^{N_\mathrm{bins}} \widetilde{N}_\mathrm{exp}^i (\widehat{\widehat{m}}_2,0,0)}\,.
\end{equation}
The definition of $h(m_2,\Delta m_{21}^2,\varphi)$ stems from \cref{eqn:ratio-MLE-g} once the Asimov dataset is considered.

\subsection{Small bin size limit}\label{subapp:Unbinned-analysis}
Once we have presented the details of the binned likelihood analysis, it is straightforward to take the limit of small bins. For a bin size $\Delta \omega_i \to 0$, the integral in \cref{eqn:Nexp-bin-i} can be computed as
\begin{equation}
\widetilde{N}_\mathrm{exp}^i(m_2,\Delta m_{21}^2,\varphi) \simeq \widetilde{n}_\mathrm{exp}(\omega_i\,; m_2,\Delta m_{21}^2,\varphi)\ \Delta\omega_i\,,
\end{equation}
where the expected number density of events per unit energy $n_\mathrm{exp}$ has been defined in \cref{eqn:nexp,eqn:nexp-tilde}. Taking this into account, and substituting the finite sum over energy bins by a continuous sum (integral) over all energies, \cref{eqn:gdis-binned-analysis} can be expressed as
\begin{equation}\label{eqn:g_dis_appendix0}
g_{a\gamma}^{\rm dis} = \sqrt{n_\sigma}\left({2\int \dd\omega\, \widetilde{n}_\mathrm{exp}(\omega\,; m_2,\Delta m_{21}^2,\varphi) \log\left[\frac{\widetilde{n}_\mathrm{exp}(\omega\,;\, m_2,\Delta m_{21}^2,\varphi)}{h(m_2,\Delta m_{21}^2,\varphi) \widetilde{n}_\mathrm{exp}(\omega\,;\, \widehat{\widehat{m}}_2,0,0)}\right]}\right)^{-1/4}\,,
\end{equation}
and 
\begin{equation}
h(m_2,\Delta m_{21}^2,\varphi) = \frac{\int \dd\omega\, \widetilde{n}_\mathrm{exp}(\omega\,; m_2,\Delta m_{21}^2,\varphi)}{\int \dd\omega\, \widetilde{n}_\mathrm{exp}(\omega\,; \widehat{\widehat{m}}_2,0,0)}\,.
\end{equation}
We can further simplify \cref{eqn:g_dis_appendix0} by using the fact that
\begin{equation}
\frac{\widetilde{n}_\mathrm{exp}(\omega\,; m_2,\Delta m_{21}^2,\varphi)}{\widetilde{n}_\mathrm{exp}(\omega\,; \widehat{\widehat{m}}_2,0,0)} = \frac{\widetilde{P}(\omega\,; m_2,\Delta m_{21}^2,\varphi)}{\widetilde{P}(\omega\,; \widehat{\widehat{m}}_2,0,0)}\,,
\end{equation}
thus leaving us with
\begin{equation}
g_{a\gamma}^{\rm dis} = \sqrt{n_\sigma}\left({2\int \dd\omega\, \widetilde{n}_\mathrm{exp}(\omega\,; m_2,\Delta m_{21}^2,\varphi) \log\left[\frac{\widetilde{P}(\omega\,;\, m_2,\Delta m_{21}^2,\varphi)}{h(m_2,\Delta m_{21}^2,\varphi) \widetilde{P}(\omega\,;\, \widehat{\widehat{m}}_2,0,0)}\right]}\right)^{-1/4}\,.
\end{equation}

\section{Exclusion limits features in the two-axion parameter space\label{app:details-exclusion-limits}}

\begin{figure*}[h]
    \includegraphics[width=0.95\textwidth]{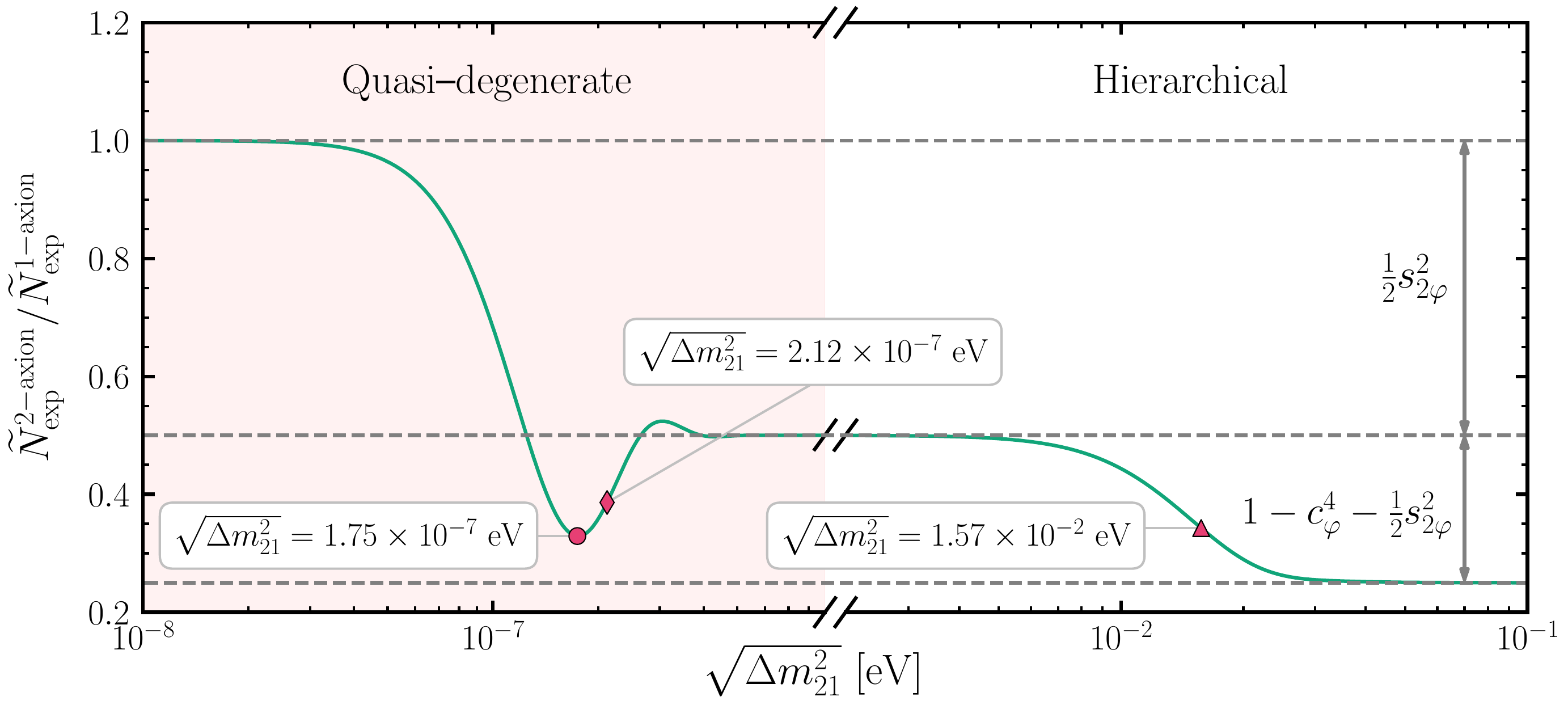}
    \caption{Expected number of photons, normalized to the single-axion prediction, in the quasi-degenerate $\sqrt{\Delta m_{21}^2} \simeq m_{1,2}$ (\emph{left}, light red) and hierarchical $m_1 \ll m_2 \simeq \sqrt{\Delta m_{21}^2}$ (\emph{right}) mass regimes for equal photon couplings $\varphi = \pi/4$. In both regimes, we are assuming the lightest axion to satisfy $m_1 < m_{c,\,\mathrm{magnet}} \sim 10^{-2}$ eV, while $m_a = 10^{-3}$~eV in the single-axion theory. \emph{Left:} The marker \protect\reddiamond\ signals the optimal discrimination point, while \scalebox{0.85}{\protect\pinkcircle}\ indicates the mass splitting with the smallest number of counts in the quasi-degenerate regime. \emph{Right:} The marker \protect\pinktriangle\ signals the optimal discrimination point in the hierarchical regime.\hfill\,}
    \label{fig:h_TotalNumberNexp}
\end{figure*}

We devote this appendix to understand with some level of detail the particular shape of CAST exclusion limits and IAXO projections in the two-axion parameter space, shown in \cref{fig:SummaryPlot,fig:quasiDegRes,fig:Hier_Res}. From \cref{eqn:exclusion-limits}, it is clear that these contours are indeed determined by the expected number of photon counts predicted within the two-axion theory, \ie\ $\widetilde{N}_\mathrm{exp}(m_2,\Delta m_{21}^2,\varphi)$, scaling as $g_{a\gamma}^\mathrm{lim} \propto \widetilde{N}_\mathrm{exp}^{-1/4}$. In \cref{fig:h_TotalNumberNexp} we present this quantity, normalized to the single massless axion prediction, as a function of the mass difference $\sqrt{\Delta m_{21}^2}$.

\begin{figure*}[t]
    \includegraphics[width=0.95\textwidth]{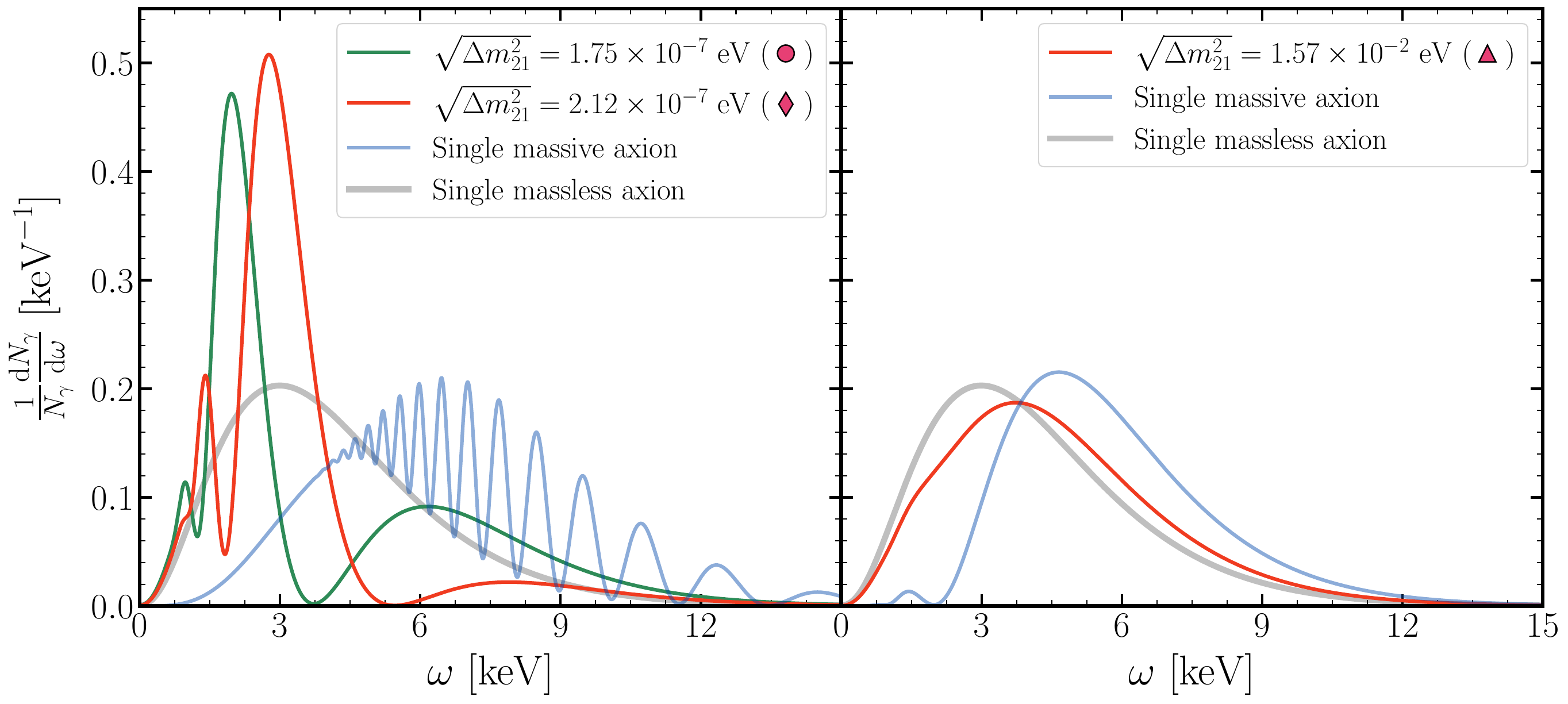}
    \caption{Normalized differential photon spectrum for quasi-degenerate (\emph{left}) and hierarchical (\emph{right}) two-axion scenarios, compared to a single massless axion (gray) and a single massive axion (blue). \emph{Left:} The green curve illustrates the spectrum corresponding to the mass difference for which the number of photon counts is minimal in the quasi-degenerate regime. The red curve, instead, corresponds to the mass difference that ensures an optimal discrimination between the two-axion and the single-axion scenarios. In both cases, we set $m_{1,2} \simeq 10^{-3}$~eV, while $m_a = 10^{-1}$~eV for the massive single-axion theory. \emph{Right:} The red curve corresponds to the mass splitting that allows an optimal discrimination in the hierarchical mass regime, which is compared with the single massive axion spectrum in blue for the same value of the axion mass. Equal photon couplings $\varphi = \pi/4$ are assumed in all cases.\hfill\,}
    \label{fig:appendix-spectra}
\end{figure*}

On the left panel of \cref{fig:h_TotalNumberNexp}, shaded light red, we illustrate the behavior within the quasi-degenerate mass regime, dictated by \cref{eqn:P-quasi-degenerate-general} under the assumption $m_{1} < m_{c,\,\mathrm{magnet}} \sim 10^{-2}$ eV, so that the oscillatory pattern in the conversion probability is entirely governed by the cosine term. It is then straightforward to check that, for $\sqrt{\Delta m_{21}^2} \ll m_{c,\,\mathrm{Sun}} \sim 10^{-7}$ eV, the oscillations are too slow to be observable and thus the expected number of photon counts approaches the single-axion prediction, \ie\ the ratio $\widetilde{N}_\mathrm{exp}^\text{2-axion}/\widetilde{N}_\mathrm{exp}^\text{1-axion} \to 1$. In the opposite limit, namely $\sqrt{\Delta m_{21}^2} \gg m_{c,\,\mathrm{Sun}}$, the cosine oscillations are too fast to be resolved and average out, reducing the number of photon counts with respect to the single-axion theory by a factor $1 - \tfrac{1}{2}s_{2\varphi}^2$, that is, $\widetilde{N}_\mathrm{exp}^\text{2-axion}/\widetilde{N}_\mathrm{exp}^\text{1-axion} \to 1/2$ for $\varphi = \pi/4$. In the intermediate range of mass differences, we highlight two important points already introduced in \cref{subsubsec:quasi_degenerate}. First, for $\sqrt{\Delta m_{21}^2} = 1.75\times 10^{-7}$~eV (marked with \pinkcircle) the number of photon counts reaches a (local) minimum. Its corresponding photon spectrum is represented by the green curve in the left panel of \cref{fig:appendix-spectra}, where one can readily see that the oscillation minimum roughly coincides with the peak of the single-axion spectrum in gray, thus producing the largest difference in the total number of counts. On the other hand, for $\sqrt{\Delta m_{21}^2} = 2.12\times 10^{-7}$~eV (\reddiamond) the discrimination between the two-axion theory and the single-axion hypothesis is optimal, in the sense that the required value of the axion-photon coupling $g_{a\gamma}^\mathrm{dis}$ is minimum (see \cref{fig:quasiDegRes}). The corresponding photon spectrum is illustrated in the left panel of \cref{fig:appendix-spectra} in red. Although the latter does not produce the largest difference in the total number of photon counts with respect to the single-axion theory, the spectral distortion is statistically more significant to discriminate between both hypotheses --- to this aim it is crucial to have enough statistics to differentiate the spectra. 

On the right panel of \cref{fig:h_TotalNumberNexp}, we turn to the hierarchical mass regime described by \cref{eqn:P-hierarchical}, assuming again $m_{1} < m_{c,\,\mathrm{magnet}}$. As discussed in \cref{subsubsec:hierarchical}, the phenomenologically relevant region within this mass regime spans around values of $m_2 \sim m_{c,\,\mathrm{magnet}}$, for which the sinc modulation becomes relevant and the cosine term averages out. Then, from \cref{eqn:P-hierarchical}, it is easy to see that for small mass differences $\sqrt{\Delta m_{21}^2} \simeq m_2 \ll m_{c,\,\mathrm{magnet}}$, the $\sinc \to 1$, and the expected number of photon counts smoothly matches the prediction obtained within the quasi-degenerate regime. On the other hand, when $\sqrt{\Delta m_{21}^2} \simeq m_2 \gg m_{c,\,\mathrm{magnet}}$, the sinc contribution is instead largely suppressed by the axion mass. As a consequence, the number of counts is suppressed by a factor $c_\varphi^4/(1-\tfrac{1}{2}s_{2\varphi}^2)$ with respect to the previous result, \eg\ $\widetilde{N}_\mathrm{exp}^\text{2-axion}/\widetilde{N}_\mathrm{exp}^\text{1-axion} \to 1/4$ for $\varphi = \pi/4$. The figure highlights the mass splitting $\sqrt{\Delta m_{21}^2} = 1.57\times 10^{-2}$~eV (\pinktriangle) for which optimal discrimination is achieved within the hierarchical regime. The associated photon spectrum is illustrated in the right panel of \cref{fig:appendix-spectra}, where it is compared with the single massive axion spectrum for the same value of the axion mass.

Finally, from \cref{fig:h_TotalNumberNexp} it is clear that the total number of photon counts within the two-axion theory is always smaller than the single-axion prediction for the same value of the coupling $g_{a\gamma}$, as it was argued in \cref{sec:axion_photon_conversion_probability_for_multiple_axions}.

\section{Axion emission from core-collapse supernovae}\label{app:SN-axions}

\begin{figure*}[t]
    \includegraphics[width=0.95\textwidth]{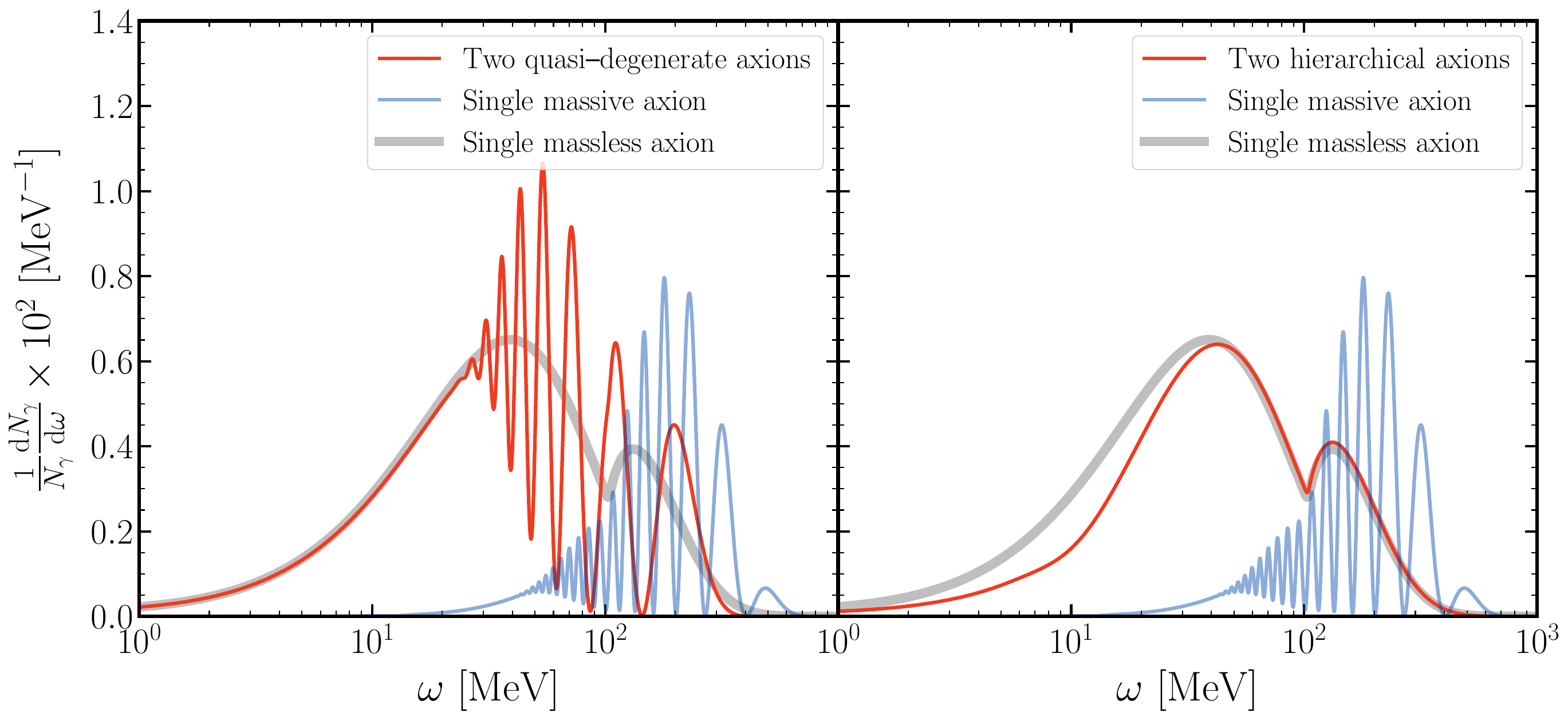}
    \caption{Normalized differential photon spectrum for quasi-degenerate (\emph{left}) and hierarchical (\emph{right}) supernova two-axion scenarios (red), compared to a single massless axion (gray) and a single massive axion (blue). Although both the single massive axion and the two-axion cases produce an oscillatory pattern that departs from the massless baseline, the spectral shapes are qualitatively different. The benchmark parameters are: $m_a = 10$~eV for the massive single-axion theory, $m_2 = 10^{-1}$~eV and $\sqrt{\Delta m_{21}^2} = 1.5\times 10^{-8}$~eV for the two-axion quasi-degenerate regime, and $\sqrt{\Delta m_{21}^2} \simeq m_2 = 1$~eV for the two-axion hierarchical regime. Equal photon couplings are assumed in all cases ($\varphi = \pi/4$).\hfill\,}
    \label{fig:SN-spectra}
\end{figure*}

Axions emitted in SN explosions are mainly produced through their coupling to nucleons $g_{aN}$ in $NN$ bremsstrahlung and pion-nucleon scattering processes. Simple expressions to parametrize the axion emission flux are provided in Ref.~\cite{Lella:2024hfk}, namely,
\begin{equation}\label{eqn:NN-brems}
\left(\frac{\dd^2 \Phi}{\dd \omega\, \dd t}\right)_{NN} = A_{NN} \left(\frac{g_{aN}}{5\times 10^{-10}}\right)^2 \left(\frac{\omega}{\omega_{NN}^0}\right)^{\beta_{NN}} e^{-(\beta_{NN}+1)\,\omega/\omega_{NN}^0}\,,
\end{equation}
for $NN$ bremsstrahlung, and
\begin{equation}\label{eqn:pionN-scattering}
\left(\frac{\dd^2 \Phi}{\dd \omega\, \dd t}\right)_{\pi N} = A_{\pi N} \left(\frac{g_{aN}}{5\times 10^{-10}}\right)^2 \left(\frac{\omega-\omega_{\pi N}}{\omega_{\pi N}^0}\right)^{\beta_{\pi N}} e^{-(\beta_{\pi N}+1)\,(\omega-\omega_{\pi N})/\omega_{\pi N}^0}\,,
\end{equation}
for pion scattering production. Summing both contributions, the total SN axion flux is given by
\begin{equation}
\frac{\dd^2 \Phi}{\dd \omega\, \dd t} = \left(\frac{\dd^2 \Phi}{\dd \omega\, \dd t}\right)_{NN} + \delta\ \left(\frac{\dd^2 \Phi}{\dd \omega\, \dd t}\right)_{\pi N}\,,
\end{equation}
where $\delta$ is a free parameter that quantifies the relative contribution of pion-nucleon scattering to SN axion production. It encodes the information on the pion abundance in the inner core region, in such a way that $\delta = 1$ corresponds to a pion-to-nucleon fraction of $Y_\pi \sim \mathcal{O}(1\%)$. As discussed in the main text, we set $\delta = 1$, although we should emphasize that there is still significant uncertainty regarding the actual pion abundance in the nuclear medium \cite{Caputo:2024oqc,Fore:2019wib}, which, in turn, limits our knowledge on the SN axion spectrum and the actual reach of the two-axion discrimination region. The previous formulae should be integrated over eight seconds after the core bounce. On that regard, the remaining parameters in \cref{eqn:NN-brems,eqn:pionN-scattering} behave as:
\begin{align}
&A_{NN} = A_{NN}^\prime\ t^{1.865}\ e^{-1.345\,t}\,,\qquad\qquad A_{NN}^\prime = 1.75\times 10^{55} \text{ MeV}^{-1} \text{ s}^{-1}\,, \\[5pt]
&\omega_{NN}^0 = \omega_{NN}^{0\, \prime} \ t^{0.755}\ e^{-0.413\,t}\,,\; \qquad\qquad \omega_{NN}^{0\, \prime} = 102.10 \text{ MeV}\,, \\[5pt]
&\beta_{NN} = \beta_{NN}^\prime\ t^{0.0410}\ e^{-0.0542\,t}\,,\qquad\quad\;\; \beta_{NN}^\prime = 1.53\,,
\end{align}
and
\begin{align}
&A_{\pi N} = A_{\pi N}^\prime\ t^{5.975}\ e^{-4.944\,t}\,,\qquad\qquad A_{\mathrm{\pi N}}^\prime = 3.88\times 10^{56} \text{ MeV}^{-1} \text{ s}^{-1}\,, \\[5pt]
&\omega_{\pi N}^0 = \omega_{\pi N}^{0\ \prime} \ t^{0.304}\ e^{-0.542\,t}\,,\; \qquad\qquad \omega_{\pi N}^{0\ \prime} = 218.59 \text{ MeV}\,, \\[5pt]
&\beta_{\pi N} = \beta_{\pi N}^\prime\ t^{-0.503}\ e^{-0.019\,t}\,,\qquad\quad\;\;\, \beta_{\pi N}^\prime = 1.27\,,\\[5pt]
&\omega_{\pi N} = \omega_{\pi N}^\prime \left(1+1.537\,t^{0.050}\right)\,, \qquad\quad \omega_{\pi N}^\prime = 40.07 \text{ MeV}\,.
\end{align}
The number of expected photon counts can be computed as
\begin{equation}
N_\gamma = \frac{S}{4\pi L_\mathrm{SN}^2} \int \dd\omega\ \varepsilon(\omega)\, \frac{\dd \Phi}{\dd \omega}(\omega)\, P_{a_\gamma\to\gamma}(\omega)\,,
\end{equation}
where $L_\mathrm{SN}$ is the distance to the SN event, $S$ the total cross-sectional area of the magnetic bore, and $\varepsilon$ the gamma-ray detector efficiency.

Taking the previous expressions at face value, in \cref{fig:SN-spectra} we illustrate the normalized differential photon spectrum from SN axion conversion in both the quasi-degenerate and hierarchical mass regimes. In the quasi-degenerate regime (left panel), the two-axion spectrum in red is modulated by the cosine oscillation around the single massless axion baseline in gray, which shows a characteristic bimodal shape that peaks at the energies where the $NN$ bremsstrahlung and pion-nucleon scattering production mechanisms become more efficient, roughly 40 MeV and 130 MeV, respectively. On the other hand, the oscillatory pattern produced by a single massive axion in blue is shifted to larger energies. Finally, in the hierarchical mass regime (right panel), the two-axion spectrum is controlled by a sinc oscillation and slightly departs from the single massless axion line.

\begin{figure}
     \includegraphics[width=0.45\columnwidth, trim={7pt 0 0 0}, clip]{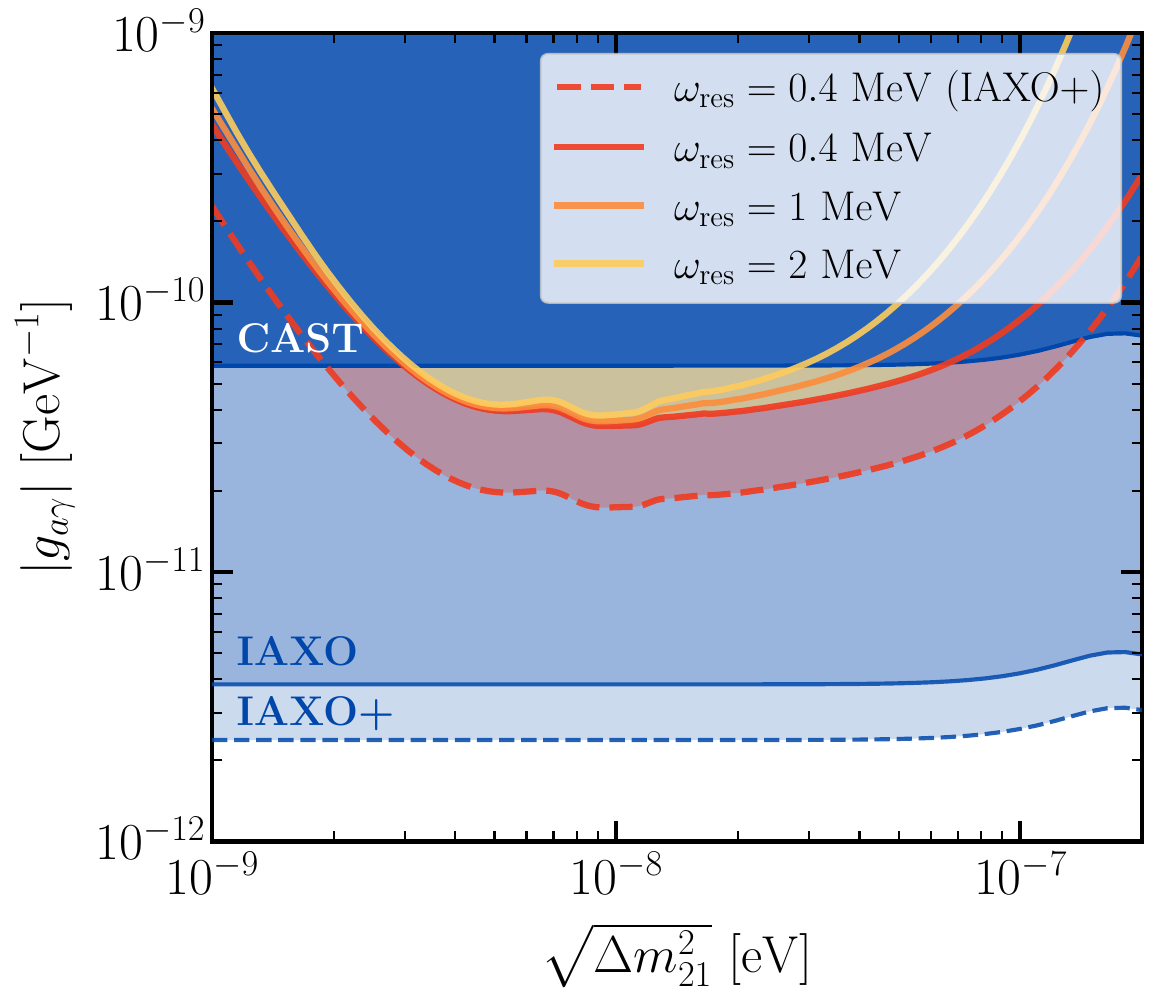}\quad 
     \includegraphics[width=0.45\columnwidth, trim={7pt 0 0 0}, clip]{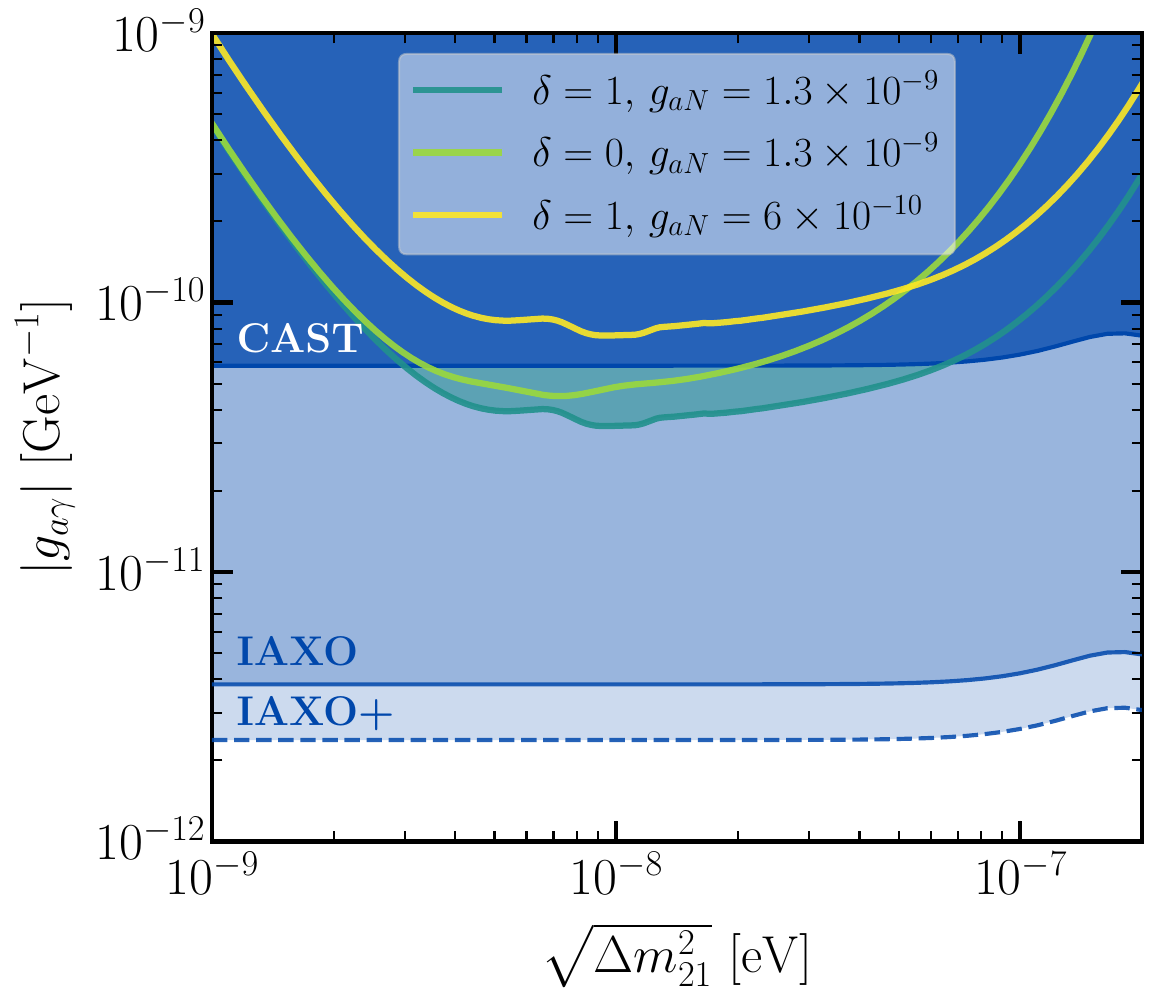}
     \caption{Two-axion discrimination reach from the analysis with SN axions in the quasi-degenerate regime $\Delta m_{21}^2 \ll m_{1,2}^2$. The axion-photon coupling is fixed to $\varphi = \pi/4$. The solid blue curve is the CAST exclusion limit derived in \cref{subsec:exclusion-limits_2-axions}, and the light blue band marks IAXO/IAXO+ projections. \emph{Left:} The yellow, orange and red lines show $g_{a\gamma}^{\rm dis}$ at $3\sigma$ for energy resolutions $\omega_{\rm res} = 0.4,\,1,\,2$~MeV, respectively (IAXO+ shown as dashed red). In all cases we set $\delta = 1$ and $g_{aN} = 1.3\times 10^{-9}$. \emph{Right:} The green, light green, and yellow lines show $g_{a\gamma}^{\rm dis}$ at $3\sigma$ for the three benchmark scenarios. In the three cases, the resolution is fixed to $\omega_{\rm res} = 0.4$~MeV. \hfill\,}
     \label{fig:SN_discovery}
\end{figure}

For completeness, in \cref{fig:SN_discovery} we illustrate in detail the discrimination region arising from the analysis of SN axions for the three benchmark scenarios with $\{\delta,\ g_{an},\ g_{ap}\}$: (i) $\{1,\ 1.3\times10^{-9},\ 0\}$, (ii) $\{0,\ 1.3\times10^{-9},\ 0\}$, and (iii) $\{1,\ 0,\ 6\times10^{-10}\}$. The left panel shows the results within the benchmark scenario (i) for different detector energy resolutions. For benchmark (ii) where the pion-nucleon scattering contribution to the SN axion flux is completely neglected the discrimination region gets reduced. For benchmark (iii) the discrimination region would already be excluded by CAST.

In this case, the distinction between the two-axion theory and the single-axion hypothesis is only possible thanks to the modulation introduced by the cosine term for mass differences $\sqrt{\Delta m_{21}^2} \sim m_{c,\,\mathrm{SN}} \sim 10^{-8}$ eV. First, individual axion masses cannot be lighter than $\sim 10^{-4}$ eV, as imposed by astrophysical constraints in the relevant range of axion-photon couplings. Second, IAXO would be insensitive to masses beyond $\sim 10^{-1}$ eV. These two conditions imply that the individual sinc terms in \cref{eqn:Paxions2photon-vac} go to unity, in such a way that the oscillations are governed by the cosine term. Since the latter is only relevant for mass splittings $\sim 10^{-8}$ eV, the two axions are quasi-degenerate.

Contrary to our analysis with solar axions, there is no available discrimination region when SN axions are hierarchical. As commented above, individual axion masses cannot be lighter than $\sim 10^{-4}$ eV according to astrophysical searches. This means that the cosine oscillations controlled by $\sqrt{\Delta m_{21}^2} \simeq m_2$ are too fast to be resolved and average out in \cref{eqn:Paxions2photon-vac}. Since we are only left with the sinc modulation, the only remaining possibility to distinguish the two-axion signal from the single-axion hypothesis would be triggered by $m_1 \ll m_2 \simeq \sqrt{\Delta m_{21}^2} \sim 1$ eV, that is, with $m_2 \sim m_{c,\,\mathrm{magnet}}$. However, this scenario requires large values of the axion-photon coupling in order to have enough statistics to make the distinction, which are already excluded by CAST. 

These features are illustrated in \cref{fig:MassRegimes}. The discrimination region for SN axions, shown in pink, is centered around $\sqrt{\Delta m_{21}^2} \sim m_{c,\,\mathrm{SN}}$ in the quasi-degenerate regime, and spans axion masses in the range $10^{-4} \text{ eV} \lesssim m_1 \lesssim 10^{-1}$ eV. As we have just argued, no analogous region arises in the hierarchical mass regime. 

\end{document}